\documentclass[traditabstract]{aa}   
\usepackage{graphicx}             
\usepackage{natbib}
\usepackage{subfigure}
\usepackage{ulem}
\usepackage{dcolumn}
\usepackage{txfonts}
\usepackage{rotating}
\renewcommand{\emph}[1]{\textit{#1}}

\include{definitions}
      \def\new#1 {{\bf #1 }}
      \def\cut#1 {\sout{#1} }

\def\kms {$\mathrm{km\,s^{-1}}$} % km s^-1
\def\HII {H{\sc ii}} % H II
\def\AMM {$\mathrm{NH_3}$} %NH3
\def\DAMM {$\mathrm{NH_2D}$} %NH2D
\def\NTH {$\mathrm{N_2H^{+}}$} %N2H+
\def\NTD {$\mathrm{N_2D^{+}}$} %N2H+
\def\HTD {$\mathrm{H_2D^{+}}$} %N2H+
\def\CO {$\mathrm{C^{18}O}$} %c18o
\def\hco {$\mathrm{HCO^{+}}$} %hco+
\def\percc {$\mathrm{cm^{-3}}$} %cm^-3
\def\cmsq  {$\hbox{{\rm cm}}^{-2}$}    %cm-2
\def\Msol {$\hbox{M}_\odot$}

\def\simgreat{\mathbin{\lower 3pt\hbox
     {$\rlap{\raise 5pt\hbox{$\char'076$}}\mathchar"7218$}}}
\def\simless{\mathbin{\lower 3pt\hbox
     {$\rlap{\raise 5pt\hbox{$\char'074$}}\mathchar"7218$}}}

\begin{document}

\title{Probing the initial conditions of high-mass star formation --
  II} \subtitle{Constraints on theory from fragmentation, stability,
  and chemistry towards high-mass star-forming regions G29.96$-$0.02
  and G35.20$-$1.74 \footnote{Fits images of PdBI continuum
    observations associated with Figs.\,1--4 are only available in
    electronic form at the CDS via anonymous ftp to
    cdsarc.u-strasbg.fr (130.79.128.5) or via
    http://cdsweb.u-strasbg.fr/cgi-bin/qcat?J/A+A/},\footnote{Based on
    observations carried out with the IRAM Plateau de Bure
    Interferometer. IRAM is supported by INSU/CNRS (France), MPG
    (Germany) and IGN (Spain).}}

  % \date{Received September 15, 1996; accepted March 16, 1997}
\author{T.\ Pillai\inst{1,2}  \and J.\
  Kauffmann\inst{2,3,4} \and F.\ Wyrowski\inst{5}   \and J.\ Hatchell \inst{6} \and A.G.\
  Gibb\inst{7} \and M.A.\ Thompson\inst{8}}

   \offprints{tpillai@astro.caltech.edu}

   \institute{ California Institute Of Technology, 1200 E California Blvd, Pasadena, CA 91125
     % \email{tpillai@cfa.harvard.edu},
    \and
  Harvard-Smithsonian Center for Astrophysics, 60 Garden
     Street, Cambridge, MA 02138
     \and
     Initiative in Innovative Computing (IIC), 60 Oxford Street,
     Cambridge, MA 02138
     \and
     Jet Propulsion Laboratory, 4800 Oak Grove Drive, Pasadena, CA 91109, USA
     \and
     Max-Planck-Institut f\"{u}r Radioastronomie, Auf dem H\"{u}gel
     69, D-53121 Bonn, Germany
     \and
     University of Exeter, UK
     \and
     Department of Physics \& Astronomy, University of British
     Columbia, Vancouver, BC, V6T 1Z1, Canada
     \and
     Centre for Astrophysics Research, Science \& Technology Research
     Institute, University of Hertfordshire, College Lane, Hatfield,
     AL10 9AB, UK
   }

 \abstract
 {Most work on high-mass star formation has focused on observations
 of young massive stars in protoclusters. Very little is
 known about the preceding stage. Here, we
 present a new high-resolution study of
 pre-protocluster regions in tracers exclusively probing the coldest
 and dense gas ($\rm NH_2D$). The two target regions \object{G29.96$-$0.02}
 and \object{G35.20$-$1.74} (\object{W48}) are drawn
 from the SCAMPS project, which searches for pre-protoclusters near
 known  ultracompact \HII\ regions. We used our data to constrain the chemical, thermal, kinematic, and
 physical conditions (i.e., densities) in G29.96e and G35.20w. \AMM, \DAMM, and continuum emission were mapped using the
   VLA, and PdBI. In particular, \DAMM\ is a unique tracer of
 cold, precluster gas at high densities, while \AMM\ traces both the cold and warm gas of modest-to-high densities. In G29.96e, Spitzer images reveal two massive
 filaments, one of them in extinction (infrared dark cloud).  Dust and line observations reveal fragmentation
 into multiple massive cores strung along filamentary structures. Most of these are cold ($< 20 ~ \rm K$),
 dense ($> 10^5 ~ \rm cm^{-3}$) and highly deuterated
 ([\DAMM/\AMM]$ > 6\%$). In particular, we observe very low line
 widths in \DAMM\ (FWHM $\lesssim 1 ~ \rm km \, s^{-1}$). These
 are very narrow lines that are unexpected towards a region forming
 massive stars.  Only one core in the center of each filament appears to be forming massive stars (identified by the presence of  masers and massive outflows);
 however, it appears that only a few such stars are  currently forming (i.e., just a single Spitzer
 source per region).  These multi-wavelength, high-resolution observations of high-mass
  pre-protocluster regions show that the target regions are
  characterized by (\emph{i})  turbulent Jeans fragmentation of massive
  clumps into cores (from a Jeans analysis); (\emph{ii}) cores and
  clumps that are ``over-bound/subvirial'', i.e.\ turbulence is too weak to
  support them against collapse, meaning that (\emph{iii}) some models
  of monolithic cloud collapse are quantitatively inconsistent with
  data; (\emph{iv}) accretion from the core onto a massive star, which
  can (for observed core sizes and velocities) be sustained by
  accretion of envelope material onto the core, suggesting that
  (similar to competitive accretion scenarios) the mass reservoir for
  star formation is not necessarily limited to the natal core;
  (\emph{v}) high deuteration ratios ([\DAMM/\AMM]$ > 6\%$), which
  make the above discoveries possible; (\emph{vi}) and the destruction of \DAMM\  toward embedded stars.}

\authorrunning{Pillai et al.\ 2010a}
\titlerunning{Fragmentation, Structure \& Chemistry in High Mass Clumps}
\maketitle

\section{Introduction} 

High-mass stars are known to form in clusters and, given their rarity,
most such clusters have typical distances of a few kpc or greater.  In
spite of being few in number relative to solar-mass stars, high-mass stars play a large role in shaping the Galaxy in each of their evolutionary stages,
with energetic stellar winds, powerful outflows, UV radiation,
expanding \HII\ regions, and finally supernova explosions. Observations
until perhaps the last decade  were hindered by a lack of
 sensitivity and resolution, crucial for such regions, which are more
 crowded and farther away than typical low-mass star-forming
 regions. 

With the development of infrared sky surveys (starting with IRAS, then
ISO, MSX, and recently Spitzer) and sensitive ground-based mm/submm
bolometer observations the potential early phases
in the formation of high-mass stars/clusters have been identified. 

Combining these results with higher angular resolution follow-up observations
with radio interferometers yields a suggested
evolutionary sequence that separates the actively accreting high-luminosity
protostars in clusters with no cm continuum (high-mass protostars
$\rightarrow$ hot molecular cores in protoclusters) from those with cm continuum
(hypercompact and ultra-compact \HII\ regions: see
\citealt{beuther2007:prpl} for a review).

The earliest stage in this scenario would be a ``pre-protocluster
stage'', which is
harder to find. The properties of this precluster stage sets the
initial conditions for the formation of high-mass stars. Observers
have attempted to identify candidate pre-protocluster cores in many
different ways (\citealt{hill2005:1.2mm},
\citealt{sridharan2005:hmsc_catalog}, \citealt{rathborne2006:irdc_bolometer},
\citealt{pillai2007}).  Follow ups with outflow tracers combined with
sensitive Spitzer observations have however revealed active star
formation in many of these cores (for e.g. \citealt{rathborne2005},
\citealt{pillai2006a:g11}, \citealt{beuther2007:protostars_irdcs}).  A
frequently used way of identifying the precluster phase has been to scan the
infrared data for dark absorption features against the bright
mid-infrared background. Several thousand such dark clouds, now
commonly known as infrared dark clouds  (IRDCs: \citealt{egan1998:irdc},
\citealt{simon2006:irdc_catalog}, \citealt{peretto2009}) have been identified along the
Galactic Plane. One of the filaments studied in this paper is such an IRDC.

To our knowledge, interferometer observations on the high-mass
precluster phase (as opposed to protocluster phase) in dense gas tracers have been few
(\citealt{swift09:irdc}, \citealt{zhang09_g28},
\citealt{busquet2010:nh2d}).  The few interferometer observations that
have been done focused on the
dust continuum structure or the chemistry. The dust continuum is
expected to be weak at these scales (as opposed to the strong dust
continuum from the envelope as observed by single dishes). This is
because precluster phase is colder and less
concentrated  than the protocluster
phase, and therefore such observations often detect only the protoclusters. Therefore, until
the commissioning of ALMA, we have to rely on molecular
observations that trace the dense and cold gas.  Ammonia and \NTH\ has
proven to be an important tool in measuring the physical conditions in
the prestellar stages \citep{tafalla2002:depletion}.  These molecules do not deplete from the gas phase for the
high densities ($<10^6$~\percc) and cold temperatures ($<20$~K)
expected in the prestellar stages. Emission from deuterated analogues (\DAMM,
and \NTD) of these two molecules is sensitive not only to dense, but also very cold gas.
Warmer temperatures
($>20$~K) usually inhibit deuterium enhancement (see
\citealt{bergin2007:araa} for a review of chemistry in dark clouds).

We embarked on a project to identify this very early stage in
the formation of a high-mass star. We identified secondary high-mass
cold cores in the close vicinity of UCH{\sc ii} regions in a SCUBA
wide-field mapping survey of $10\times 10$~\arcmin\ (SCAMPS: the
\underline{SC}UB\underline{A} \underline{M}assive
\underline{P}re/Protocluster core \underline{S}urvey; \citealt{thompson2005:dmuconf}). Recently,
we reported on a single dish study of \AMM\ deuteration
([\DAMM/\AMM]) and depletion (of CO) toward a sub-sample of these new
massive pre/protocluster cores. More than half of the sources exhibit
a high degree of deuteration ($0.1 - 0.7$).  The enhancement in
deuteration coincides with moderate CO depletion onto dust grains
\citep{pillai2007}. Here, we present interferometer observations of
\AMM\ and \DAMM\ using the VLA, BIMA and PdBI for two sources from the
SCAMPS project.

The SCAMPS source which we denote \object{G29.96e} in this paper, is an IRDC about two arcmin east of the ultracompact H{\sc ii}
 region G29.96-0.02. The cometary H{\sc ii} region
itself is one of the brightest radio and infrared sources in our
Galaxy, and has been studied in detail at infrared and radio
wavelengths (\citealt{wood1989:uchii}, \citealt{pratap1999}; \citealt{kirk2010}). We adopt
the kinematic distance of 7.4~kpc to the cloud estimated from the
\DAMM\ LSR velocity.

The other SCAMPS source which we denote \object{G35.20w} lies about one arcmin west
of the W48 H{\sc ii} region (G35.20$-$1.74). We adopt a distance of 3.27~kpc based on recent trigonometric parallax measurements by
\citet{zhang09_g35}. This value is very close to our
kinematic distance estimate of 3.4~kpc for the region. The position of
the high-mass cloud we found with SCUBA roughly coincides with that of
a molecular cloud adjacent to W48 region mapped at coarse resolution
in CO, and its isotopomers
(\citealt{vallee1990}, \citealt{zeilik1978}). This region has been studied in dust polarization by 
\citealt{curran2004}, where our G35.20w is referred to as W48W.

By determining the properties of these massive secondary cores from
our high angular resolution observations, we will constrain the
initial conditions of high-mass star formation. After presenting the
observations (Section \ref{sec:obs}), we will discuss our main results
(Section \ref{sec:res}),  then analyze these results (Section
  \ref{sec:analysis}), and finally discuss the relevance of these
  results in the context of current theories on initial fragmentation
and collapse in high-mass star-forming regions
(Section \ref{sec:theory}).

%%%%%%%%%%%%%%%% ############ OBSERVATION ############## %%%%%%%%%%%%%%%%%%%%%%%%
%%%%%%%%%%%%%%%% ############ OBSERVATION ############## %%%%%%%%%%%%%%%%%%%%%%%%
%%%%%%%%%%%%%%%% ############ OBSERVATION ############## %%%%%%%%%%%%%%%%%%%%%%%%

\begin{figure*} \centering
\includegraphics[height=\linewidth,angle=-90]{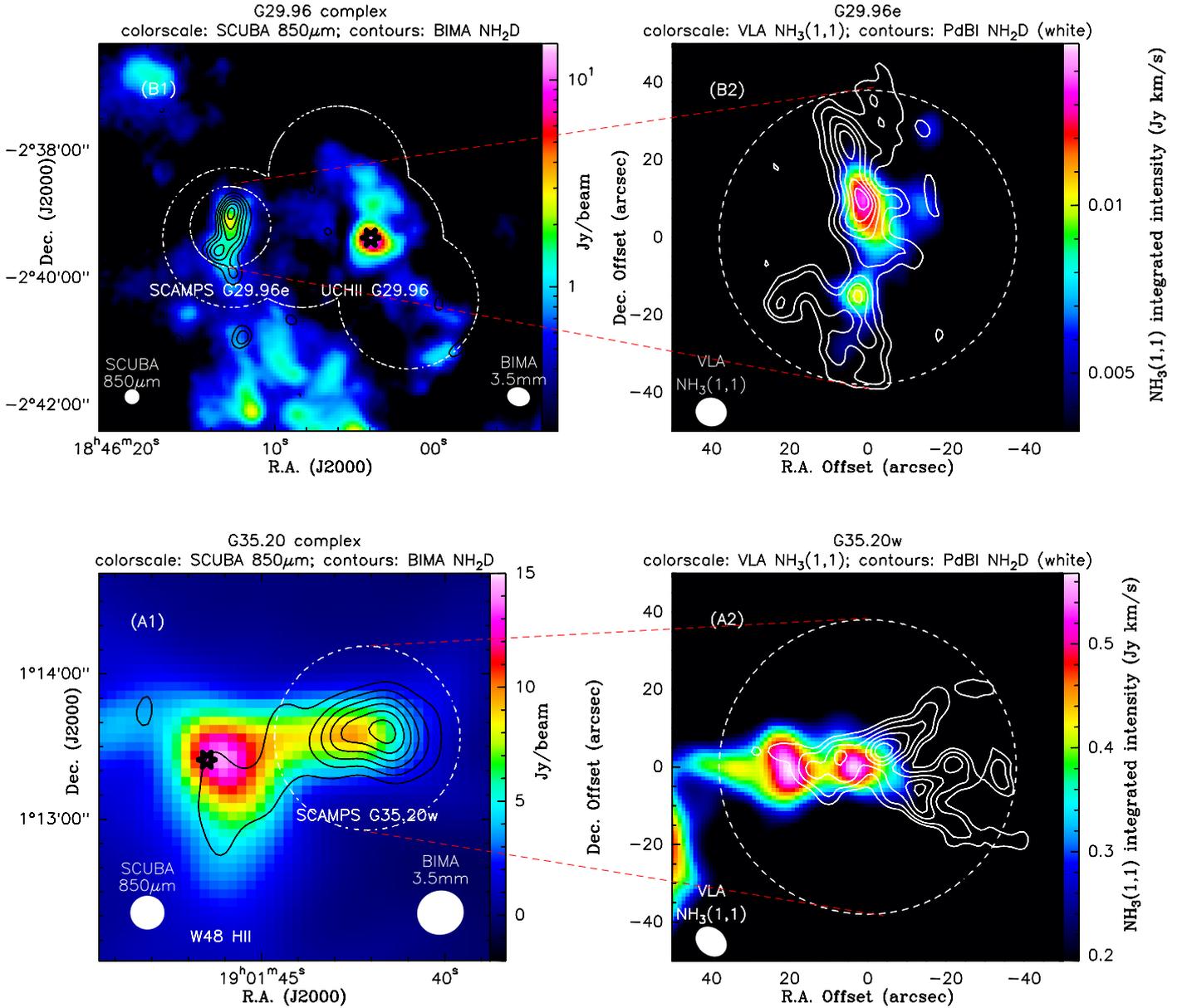}
\caption{Overview of the SCAMPS regions, G29.96e (\textit{top}) and
  G35.20w (\textit{bottom}).  \textit{Panel (A1)}:  SCUBA
    850$\mu$m dust continuum (color scale)
   and BIMA  \DAMM\ at 85.9~GHz (contours).\textit{Panel (A2)}: VLA
   \AMM\ (1,1) (color scale) starting at 3$\sigma$ and PdB \DAMM\ at 85.9~GHz (white
   contours). \textit{Panel (B1)}: SCUBA 850$\mu$m dust continuum (color scale) and BIMA \DAMM\ at 85.9~GHz  (contours). The white dash circle indicates the BIMA coverage. \textit{Panel (B2)}: VLA
   \AMM\ (1,1) integrated intensity (color scale) starting at 3$\sigma$ and PdB \DAMM\ at 85.9~GHz integrated intensity (white
   contours).  The
   contour levels for PdBI maps start at
   $-3\sigma$, $3\sigma$ in steps of 3$\sigma$.   The primary beam at  3.5~mm is indicated by white dashed circles in the right panels.  The nominal positions of the H{\sc ii} regions are marked as stars \citep{wood1989:uchii}.}
\label{fig:overview}
\end{figure*}

\section{Observations \label{sec:obs}}

Here we describe the radio interferometer observations with
various telescopes, in line and continuum, as well as the Spitzer
archival data. All the figures shown in the paper (except for the
first reference figure, figure~1) are with respect to the phase center
of the Plateau de Bure Interferometer (PdBI) observations
(Table~\ref{tab:source_list}). A summary of line and continuum
frequencies relevant to this paper is given in
Table\ \ref{tab:freq_list}. The final synthesized beam, as well as the
root mean square (RMS) noise level for each set of observations is given in Table\
\ref{tab:beam_rms}.

%\sc for BOLD
% TABLE 1. SOURCE LIST
\begin{table}
\caption{Reference Position for PdBI Interferometer Observations}
\vspace{1em}
\begin{tabular}{rccc}
\hline
{Source} & { R.A.(J2000)}& {Dec.(J2000)}    & { $v_{\rm LSR}$ [km/s]}   \\
\hline
G29.96e      &     18:46:12.78     & $-$02:39:11.8  & 100.4     \\
G35.20w      &     19:01:42.11     & $+$01:13:33.4  & 42.4      \\
\hline
\end{tabular}
\label{tab:source_list}
\end{table}

The SCUBA observations have  been reported in \citealt{thompson2005:dmuconf}, and will be detailed in a separate paper (Thompson et al. in preparation).

%Table 1.  PARAMETER TABLE FOR OBSERVED ROTATIONAL TRANSITIONS
\begin{table}[h]
\begin{center}
\caption{Relevant observed rotational transitions.}
\vspace{1em}

\begin{tabular}{lcccc}
\hline
 Species & Transition       & Array & $\nu$ (MHz) & \\
\hline
\DAMM\ & 1$_{11}$--1$_{01}$ & PdBI &  85926.3      \\
\AMM\  & (1,1)              & VLA & 23694.496    \\
\AMM\  & (2,2)              & VLA & 23722.633    \\
\hco   & 1--0               & BIMA & 89188.5230 \\
\hline
\end{tabular}
\end{center}
\label{tab:freq_list}
\end{table}

\subsection{PdBI  observations}
G29.96e and G35.20w were observed with the PdBI in its C
and D configurations on Mar 27/28th and Apr 20/21st 2004.  The 3.5\,mm
receivers were tuned to the $\rm NH_2D$ line at 86.086 GHz in single
sideband mode.  The 1.3\,mm receivers were tuned to 219.560 GHz in DSB mode. The 3.5\,mm lines were observed with a resolution of 80 kHz. At both wavelengths two 320 MHz wide correlator windows were used to probe the continuum emission. The spectral
resolution for the \DAMM\ observations is $\approx 0.27$~\kms.

Hexagonal mosaics with a spacing of 10 arcsec were used to increase
the field of view. This kept the sensitivity at the mosaic center very
high for the 3.5\,mm observations and led to fully sampled mosaics at
1.3\,mm. 

Using natural weighting, the synthesized beam sizes, and sensitivities
at 3 and 1.3mm are as given in Table~\ref{tab:beam_rms}.

\subsection{VLA observations}
The 23~GHz \AMM\ observations are reported for two sources G35.20w and
G29.96e.  We retrieved the data for \AMM\ (1,1) and (2,2) transitions at 1.3\,cm
from the VLA archive\footnote{ The National Radio Astronomy
Observatory is a facility of the National Science Foundation operated
under cooperative agreement by Associated Universities, Inc.} toward
G35.20w, while for G29.96e, we report new observations with the VLA.

The 23~GHz G29.96e \AMM\ observations were done with the Very Large Array
(VLA) on 24 August 2004 in its D configuration and in 2 polarizations
with a spectral resolution of 0.33~\kms. The primary beam at 23~GHz
is $\sim$2~arcmin. The observations were carried out in
a three pointing mosaic covering the SCAMPS source G29.96e as well
as the cometary H{\sc ii} region G29.96$-$0.02. The standard interferometer mode was
used with a total integration time of 20 minutes on source split into
sessions.  Phase calibration was done using J1733-130 and the flux
calibrator was 3C286. The archival data retrieved for G35.20w
  were  obtained in D array on 03 August 2000, covering the (1,1)
  and (2,2) lines with 2$\times$64 49~kHz channels. The phase center was
  at $\alpha_{\rm J2000}=$19:01:45.5, $\delta_{\rm J2000}=$$+$01:13:28.00. The total integration time on source was approximately 50 minutes. The phase calibrator J1824+107 was observed every 10 minutes and the flux calibrator was 3C286.

\subsection{BIMA observations}
G35.20w and G29.96e were observed with BIMA in its D configuration in
Jul/Aug. 2003 in three tracks. The receivers were tuned to the $\rm
NH_2D$ frequency. Simultaneously, HCO$+$ (1-0) was observed 
in the upper sideband. The frequency resolution was 0.1
MHz. More details on the observations are as given in Table~\ref{tab:beam_rms}. Towards G29.96e, a 5 point mosaic was observed to cover both
the hot core and the SCAMPS source in the east. Phase calibration with 1743-038 was performed every 15 minutes, and Uranus was observed as flux calibrator.

%TABLE 3.Interferometer Observations
\begin{table*}[p{0.1cm}]
\begin{center}
\caption{   The sensitivity and beam parameters for the interferometer observations.  Columns are the interferometer used, the wavelength with
the molecule in brackets for line observations, the effective beam, its position angle and the RMS noise level in the final image.}
\small
\begin{tabular}{lcccc}
\hline
 Instrument            & Wavelength(type)   & $\theta_{\rm maj} \times \theta_{\rm min}$ & PA        & RMS (type)     \\  
                       &                    &    arcsec                                  & degrees    & mJy/beam  \\    
\hline
                       &                    &    {\bf G29.96e}                                &            &           \\
 PdBI                  & 1.3mm                &   2.2   $\times$  1.6                      & 44       & 2.3 (continuum)      \\
                       & 3.5mm (\DAMM)              &   5.5   $\times$  4.1   & 43       & 0.5 (continuum), 35 (line) \\
 VLA                   & 23.6~GHz  (\AMM)            &   8.0 $\times$  7.4  & 12 &  1.1 (line)   \\
BIMA                   & 89.2~GHz (HCO$+$ 1--0) & 21.7 $\times$ 17.6 & 14.3 & 420 (line) \\
                       &                    &    {\bf G35.20w  }                                 &             &           \\
 PdBI                  & 1.3mm               &   1.8   $\times$  1.7                     &  52       & 2.5 (continuum)       \\
                       & 3.5mm  (\DAMM)               &   4.5   $\times$  4.2                     &  45       & 0.5(continuum), 31 (line)       \\
 VLA                   & 23.6~GHz  (\AMM)            &   8.7   $\times$  6.9 &  42 & 1.2 (line) \\
BIMA                   & 89.2~GHz (HCO$+$ 1--0) & 19.2 $\times$ 18.2 & -20.6 & 210 (line)\\
\hline
\label{tab:beam_rms}
\end{tabular}
\end{center}
\end{table*}

\subsection{Spitzer archival data \label{sec:obs_mir}}
The Spitzer MIR data collected from the Spitzer data archive is part
of the GLIMPSE (\citealt{benjamin2003,churchwell2009}, and MIPSGAL Legacy survey (\citealt{carey2009}) in 4 of the Infrared Array
Camera (IRAC) bands from $3.6$ to 8$ \mu$m, and two of the Multiband
Imaging Photometer (MIPS) bands at 24, and 70$\mu$m, respectively,
These two Galactic Plane surveys cover the inner Galactic Plane, and
thus include the G29.96e region but not
G35.20w which lies outside the latitude limits of GLIMPSE \& MIPSGAL. However, we were able to download the MIPS 24$\mu$m data for
G35.20w from a dedicated survey of UCHII regions by Carey et al (Spitzer Program ID 20778). Our
molecular cloud happened to be just within the 24$\mu$m field of view
of their target, but the 70$\mu$m image did not cover this cloud.

We discuss  G29.96e, followed by G35.20w in each subsection below.

%%%%%%%%%%%%%%%% ############ RESULTS ############## %%%%%%%%%%%%%%%%%%%%%%%%
%%%%%%%%%%%%%%%% ############ RESULTS ############## %%%%%%%%%%%%%%%%%%%%%%%%
%%%%%%%%%%%%%%%% ############ RESULTS ############## %%%%%%%%%%%%%%%%%%%%%%%%

\section{Results\label{sec:res}}

\subsection{Methods}
\subsubsection{Identification of \AMM\ and \DAMM\ clumps\label{sec:method_line_clump} }
We used CLUMPFIND, a 3-dimensional clump finding algorithm, to identify
the \DAMM\ cores \citep{williams1994_clfind}. The threshold for \DAMM\
was set to $0.2 ~ \rm Jy \, beam^{-1}$ for both sources. The step size was
$0.02 ~ \rm Jy \, beam^{-1}$, where the RMS noise level (measured in
regions free of emission) is $35 ~ \rm mJy \, beam^{-1}$. For \AMM\ observations of G29.96e, the threshold was set to
$0.005 ~ \rm Jy \, beam^{-1}$, and we used an increment of
$0.001 ~ \rm  Jy \, beam^{-1}$. For G35.20w in this line, a higher
threshold of $0.02 ~ \rm Jy \, beam^{-1}$ was used, while the step size
was lowered to an unusually low value (a fraction of the threshold
value), in order to separate two equal intensity peaks. The observed transition
 with six hyperfines can be grouped into three pairs: a main pair
 hfs$_{\rm mg}$(which are blended), an inner pair hfs$_{\rm in}$, and
 an outer pair hfs$_{\rm out}$. We ignored the outer hyperfines and
 considered the whole velocity range within the inner hyperfines in
 CLUMPFIND, since in LTE  hfs$_{\rm out}$ are only 30\% as bright as the main
 group. We identified 11 clumps in G29.96e and 13 in G35.20w. The clumps are
 referred to as P1...P'n' (n=11 for G29.96e, and n=13 for G35.20w) in
 Table~\ref{tab:nh2d_line_parameter}. Similarly, we identified 5 and 3 \AMM\ clumps in G29.96e and G35.20w respectively. The line parameters for \AMM\ clumps are listed in Table~\ref{tab:deutn_ratio}.

\subsubsection{Identification of  PdBI dust continuum cores\label{sec:method_cont}}
A modified version of CLUMPFIND for 2-dimensional data is available.
However, on manually checking the results, we found that the structures
were not intuitively meaningful in some cases. Therefore, we chose to
identify dust emission cores as dust emission peaks of an intensity
that exceeds the noise level by at least a factor 5 and with sizes larger than a beam. We assigned
masses to them according to the following scheme. For a given core
(i.e., peak), we identify the $3~\sigma$ contour that contains this
core. If this contour contains a single peak, then all of the area
contained within the $3~\sigma$ contour is assigned to that core. If
several cores are present within the contour, then the core boundaries
are drawn at the saddle points separating these cores, respectively at
the $3~\sigma$ contour enveloping the peaks. The dust continuum mass is then computed over an
effective area of radius R$_{\rm eff}$ after a beam deconvolution. Results are presented in
Table~\ref{tab:pdbi_cont_mass}.

\subsubsection{Analysis of \AMM\ and \DAMM\  spectra \label{sec:method_line_anal}}
To determine the gas temperature and column density, we followed the
analysis routine described in \citet{pillai2007}.  The ratio of the
brightness temperatures of the \AMM\ (1,1) and (2,2) transitions,
along with the optical depth, can be used to estimate the rotational
temperature. For temperatures $< 20 ~ \rm K$, which are typical of
cold dark clouds, the rotational temperature closely follows the gas
kinetic temperature \citep{walmsley1983:nh3}. Like \AMM (1,1) and
  (2,2), its
isotopologue \DAMM\ 1$_{11}$--1$_{01}$ also has hyperfines. This allows the estimation of
optical depth and hence the column density of \DAMM.  We estimate the \AMM\ column density assuming that the level populations are
thermalized, i.e., $T_{\rm ex} = T_{\rm rot}$. A similar assumption holds for our
estimation of \DAMM\ column densities.  i.e., Assuming that \DAMM\ and
\AMM\ reside in the same volume, the excitation of \DAMM\ can be
gauged using the gas temperatures derived from the \AMM\ analysis.
This permits to derive \DAMM\ column densities. The expressions used
to estimate the \DAMM\ column
density from the radiative transfer equation is given in Appendix A of
\citet{pillai2007}. We derive the \DAMM\ column density
solely based on the ortho transition. For this, we assume that the ortho
and para transitions are in LTE. The \DAMM\ partition function is determined by considering
the contribution of the different energy levels from $J = 0$ to
$J = 2$. Then, the \DAMM\ and
\AMM\ fractionation ratio can be calculated from ratios of column
densities. We calculate the line parameters (line width, peak brightness, LSR
velocity) for both tracers for the same beam
size, averaged over a square aperture of size equal to the beam major
axis. We then estimate the deuteration ratio [\DAMM/\AMM] wherever both tracers are detected with a
signal that exceeds the noise level by a factor 3. We require that the uncertainties in their ratios are below 50\%. Table~\ref{tab:deutn_ratio} lists our results.

\subsubsection{Mass determination\label{subsec:mass_method}}
We derive the gas mass from the 1.3~mm and 3.5~mm dust
continuum emission for the different clumps identified at both
wavelengths following \citet{kauffmann2008}, using dust opacities at
1.3~mm (0.009 $\rm cm^2/g$) and 3.5~mm opacities (0.002 $\rm cm^2/g$)
for dust grains with thick ice mantles and gas density $\rm n(H)=
10^6~$\percc as in \citealt{ossenkopf1994:opacities}, and gas to dust
ratio of 100. A dust opacity index of 1.61 is obtained by interpolating
between two nearest wavelengths 1.075~mm and 1.300~mm from \citet{ossenkopf1994:opacities}. Here, we adopt dust temperatures identical to the
\AMM{}-derived ones (Table~\ref{tab:deutn_ratio}). If no such
measurement exists, a temperature of $16 ~ \rm K$ is used. This is the
case for e.g. when the \AMM\ temperature measurements have a large
($>$30\%) error.  We adopt $16 ~ \rm K$ since it is the average
temperature of the cores (Table~\ref{tab:deutn_ratio}). For a
temperature of 16~K, the 1.3~mm continuum sensitivity limit quoted in
Table~\ref{tab:beam_rms} correspond to a mass limit of $\sim 4$ and $1
\, M_{\sun} \, \rm beam^{-1}$ for G29.96e and G35.20w, respectively.

\subsection{Overall cloud structure and associated star formation \label{sec:res_structure}}
\subsubsection{G29.96e \label{subsec:res_g29structure}}
We discovered the molecular clump G29.96e in our SCAMPS
$10\arcmin \times 10\arcmin$ wide-field mapping survey of the
G29.96-0.02 region as the brightest dust clump that has no
associated $8 ~ \rm \mu m$ infrared emission (as seen with MSX). The
SCUBA map exposes G29.96e as an elongated feature that is dominated by a
single bright peak (Fig.\ \ref{fig:overview}). The estimated total
mass of this region, estimated from SCUBA data, is
$7000 \, M_{\sun}$  (Sect. \ref{subsec:mass}). The PdBI map at
$3 ~ \rm mm$ wavelength shows a filament of about $1\arcmin$
($\sim 2 ~ \rm pc$) length that runs north-south. CLUMPFIND breaks this
emission up into 3 cores. At even higher spatial resolution, the
$1 ~ \rm mm$ PdBI map reveals seven cores with a mass of $58 ~ {\rm
  to} ~ 320 \, M_{\sun}$ and a radius $\sim 0.1 ~ \rm pc$
(Sect. \ref{subsec:mass}). We denote these cores as mm1 to mm7 (the cores from the
$3 ~ \rm mm$ remain unnamed), in order of decreasing peak intensity
(Table \ref{tab:pdbi_cont_mass}).

The Spitzer $8 ~ \rm \mu m$ image of G29.96e, shown in Fig.~\ref{fig:g29_g35-all},
 reveals this region as an Infrared Dark Cloud (IRDC); it is seen in
extinction against the Galactic background. The extinction features
correlate strongly with the PdBI continuum emission. Such
extinction structures are still visible in Spitzer images at
$24 ~ \rm \mu m$ wavelength (also shown in Fig.~\ref{fig:nh2dspec}). The latter images also expose
a point source in this region, coincident with the dust core  G29.96e mm1. This
source drives an outflow at a velocity similar to one of the G29.96e
clumps (see below). This kinematic information suggests that the
$24 ~ \rm \mu m$ point source is physically associated with the G29.96e
region (i.e., does not reside in the foreground or background). The
core G29.96e mm1 is therefore apparently actively forming stars or clusters.
The Spitzer $8 ~ \rm \mu m$ images arguably show an elongated bright
structure towards G29.96e mm1; this may be due to emission from an outflow
(e.g.\ \citealt{shepherd2007_g34}). No star-formation activity is
detected in other parts of the cloud. \citet{walsh1998:meth_radio} have reported 
a Class~II methanol maser detection at 6.7~GHz towards  G29.96e ($\sim
4$~\arcsec north offset from G29.96e mm1). These masers are exclusive signposts of the earliest stages of high-mass star formation \citep{pestalozzi08:stat}.

The wide-field map from BIMA shows that \DAMM\ is only detected towards
the G29.96e region, and not towards the neighboring G29.96-0.02 H{\sc ii}
region (see Fig.\ref{fig:overview}). This is consistent with the picture of G29.96e being colder than clump that hosts the H{\sc ii} region: high temperatures are
expected to yield low abundances of \DAMM\ (Sect.~\ref{analysis:deutn-theory}). The PdBI maps
probe the \DAMM\ distribution at higher spatial resolution.
Segmentation of the map with CLUMPFIND yields 11 cores. There is a very good correlation between the PdBI-observed \DAMM\ and dust emission
distribution, as well as the Spitzer $8 ~ \rm \mu m$ extinction
structures. Details of this spatial correlation are examined in Sect.\ref{analysis:deutn}. We also smooth the PdBI \DAMM\ maps to the resolution of the VLA
\AMM\ data, in order to compare their distribution. This reveals a
strong correlation between \DAMM\ and \AMM; however, we do also find
two \AMM\ cores without \DAMM\ emission (for e.g. those at $[+5, -29]$ and
$[-16, +28]$). The \DAMM\ emission is well within the VLA
primary beam (2~\arcmin) at 23~GHz. Such details are also further examined in Sect.\ref{analysis:deutn}. The
\AMM\ analysis yields rotational temperatures (which serve as an
estimate of the kinematic temperatures) of $13 ~ {\rm to} ~ 23 ~ \rm K$.

BIMA maps in the \hco\ (1--0) line reveal emission from one or several
outflows. This is shown in Fig.\ref{fig:outflow}, where we present integrated
intensity maps for the blueshifted (91--96.0~\kms) and the
redshifted (from 108--112~\kms) outflow emission. Both
velocity-integrated signals peak at opposite sides of G29.96e mm1, which
suggests that the outflow source resides in this core. Broad line
wings are also observed with the SMA in the CO (3--2) transition.  A
detailed characterization of the outflows and the massive protocluster
in G29.96e mm1 will be presented in a future paper.

\subsubsection{G35.20w}
The clump G35.20w was found in our SCAMPS SCUBA survey map of the W48
H{\sc ii}, where it manifests as a dust filament with east-west
orientation that points away from the H{\sc ii} region (Fig.\ref{fig:overview}). Another filament with north-south orientation is detected closer
to W48; this structure is not further examined here. The G35.20w filament
has a physical size of $\sim 2 ~ \rm pc$, slightly shorter than the length of the filament in
G29.96e as observed with SCUBA. The SCUBA map reveals a filamentary structure in G35.20w: the field studied here
contains four peaks. The total mass of the filament  from the SCUBA data is $\sim 5000 \, M_{\sun}$ (see Sect.\ref{subsec:mass} for aperture and core masses). The 3.5~mm PdBI dust emission map essentially recovers the
filamentary structure already seen with SCUBA, but shows it at higher
spatial resolution. The 1.3~mm dust map from the PdBI, however, only
reveals 4 compact peaks in the map, i.e.\ the global filamentary
structure is lost because of interferometer filtering. Compared to
G29.96e, this filtering is stronger in G35.20w, since the latter source is
closer and thus larger when projected onto the sky. The compact 1.3~mm
dust emission cores, numbered mm1 to mm4, have a radius
$\sim 0.02 ~ \rm pc$ and a mass of $8 ~ {\rm to} ~ 18 \, M_{\sun}$.
This is much denser and the radii and masses  smaller than in G29.96e. Different
physical domains are thus probed in G29.96e and G35.20w.

Unfortunately, neither GLIMPSE nor MIPSGAL data are currently available for G35.20w, being at a Galactic latitude above the range covered by either surveys. As mentioned in Sect.~\ref{sec:obs_mir}, we were able
to download the MIPS 24\,$\mu$m data for G35.20w from a dedicated survey of
UC\HII\ regions by Carey et al (Spitzer Program ID 20778). Our molecular cloud happened to be just
within the 24\,$\mu$m field of view of their target, but the 70\,$\mu$m image
did not cover this cloud. The  SCUBA clump (which is resolved into ``streamers'' at higher resolution) appears to originate right at the edge of the strong MIR emission of the W48 H{\sc ii} region.
 Our brightest mm continuum source, G35.20w mm1, is associated with bright
point-like emission at 24$\mu$m. A comparison of Spitzer and \DAMM\ in
Sect.~\ref{subsec:nh2d_vs_dust} shows  no MIR emission associated with
\DAMM\ in G35.20w. 6.7~GHz Class~II methanol masers have been detected
significantly offset from the streamer, closer to the H{\sc ii} region (Goedhart, Wyrowski
\& Gibb in preparation). These masers are exclusive
signposts of the earliest stages of high-mass star formation
\citep{pestalozzi08:stat}. However unlike in G29.96e, the gas
streamer seen in \DAMM\, and dust continuum is not seen in absorption
(nor in emission) against the MIR background. This is likely because
the star-forming region itself is slightly below the Galactic Plane
which usually provides the bright MIR background which the cold dust
would absorb, and be seen as IRDCs. 

In the BIMA wide-field mapping, \DAMM\ is clearly detected in G35.20w,
but only a weak detection is found towards the north-south dust
filament closer to W48 (Fig.\ref{fig:overview}).
Given our present understanding of deuteration fractionation (Sect.\ref{analysis:deutn}), G35.20w is thus supposedly less evolved than the other filament.
The PdBI data reveal several \DAMM\ filaments.
CLUMPFIND segments these into 12 cores. The \DAMM\ emission is
strongly correlated with the filamentary dust emission, as probed at 3.5~mm wavelength. This correlation is, however, weaker than in G29.96e: it
is actually rather poor in the eastern part of G35.20w. In that region, a
weak correlation is also observed between \DAMM\ and the
VLA-probed \AMM\ emission. Conversely, the western part of G35.20w hosts
\DAMM\ cores not seen in \AMM{} map. This is not surprising because
these are the only \DAMM\
cores that lie outside the primary beam of the VLA.  Such details of the \DAMM\ data
are also examined in Sect.\ref{analysis:deutn}. The \AMM\ emission, in contrast, correlates
well with the 3.5mm dust emission. We derive rotational temperatures of
$15 ~ {\rm to} ~ 21 ~ \rm K$ from analysis of the \AMM\ emission.

BIMA \hco\ (1--0) maps reveal an outflow in G35.20w, just as observed in
G29.96e. The distribution of the velocity-integrated blueshifted
(25.5 -- 35.0~\kms) and redshifted (46.6 -- 53.6~\kms) emission is
presented in Fig.\ref{fig:outflow}. The core G35.20w mm1 is located in between the
intensity maxima of these two velocity maxima, and so it is likely the
source of the outflow. This core is thus actively forming a
star or cluster. Detailed studies of this object are not the focus of
the present paper, though. Signs for active star formation are
detected nowhere else in the field studied here.

%TABLE 5.  LINE PARAMETER TABLE FOR PdBI data
\begin{table*}[h]
\begin{center}
\caption{Line parameters, and virial mass for \DAMM\ cores.\label{tab:nh2d_line_parameter}}
\vspace{1em}
\begin{tabular}{llcccccccc}
\hline
Core & Offsets          & $v_{\rm LSR}$ & $\Delta$v$^{\rm a}$ &  $\tau_{\rm tot}$$^{\rm b}$ & R$_{\rm eff}$   &  M$_{vir}$$^{\rm c}$ & M$_{dust}$$^{\rm c}$  & $\alpha$$^{\rm d}$ & $\rho$$^{\rm e}$  \\

 &   arcsec             &  \kms\ &  \kms\           &                            &  parsec        &  \Msol\             &  \Msol\  &    & 10$^5$(\percc) \\
\hline

 &                 &                    &    {\bf  G29.96e }             &       & &   & \\
 P1 & (  6.0, 25.0)           & 103.67 (0.01) &    0.94    (0.02)   &     2.40    (0.29)    &0.24    & 44      & 255     & 0.2    &  1  \\
 P2 & (  0.0,  8.0)           & 103.17 (0.01) &   1.41    (0.02)   &     3.81    (0.2 )    &0.219   & 91      & 1168    & 0.1    &  4  \\
 P3 & ( 23.0,-16.0)$^{\rm 1}$ & 102.77 (0.02) &   0.73    (0.03)   &     2.69    (0.59)    & 0.135   & 15      & $<56$      & --     &  -- \\
 P4 & ( -2.0,-33.0)$^{\rm 1}$ & 102.87 (0.03) &   1.11    (0.04)   &     5.78    (0.78)    & 0.145      & 37     & $<9$       & --     &  -- \\
 P5 & ( 10.0,-18.0)$^{\rm 1}$ & 101.57 (0.01) &   0.72    (0.02)   &     3.26    (0.48)    & 0.127     & 12      & $<73$      & --     &  -- \\
 P6 & ( 16.0,-11.0)$^{\rm 1}$ & 102.47 (0.01) &   0.68    (0.03)   &     1.5     (0.52)    & 0.117      & 11      & $<70$      & --     &  -- \\
 P7 & ( -1.0,-26.0)$^{\rm 1}$ & 103.27 (0.02) &   0.98    (0.04)   &     7.48    (0.9 )    & 0.125      & 25      & $<75$      & --     &  -- \\
 P8 & (  3.0,-13.0)           & 103.37 (0.03) &   1.22    (0.04)   &     5.0     (0.56)    &0.074   & 23      & 132     & 0.2    &  11 \\
 P9 & (  3.0,  6.0)           & 103.17 (0.01) &   1.42    (0.02)   &     3.74    (0.2 )    &0.19    & 80      & 986     & 0.1    &  5  \\
 P10 & ( -6.0, -4.0)          & 102.37 (0.27) &   0.8     (0.91)   &     0.95    (0.1 )    &0.154   & 21      & 289     & 0.1    &  3  \\
 P11 & (-15.0, 28.0)$^{\rm 3}$& 101.87 (0.03) &   0.69    (0.05)   &     7.29    (1.95)    &--      & --      & $<7$       & --     &  -- \\
  \hline
 &                 &      &              &  {\bf  G35.20w}               &       & &    &  \\
 \hline
 &              &    \kms\   & \kms\           &                            &  parsec        &  \Msol\             &  \Msol\  &    & 10$^5$(\percc) \\
 \hline
 P1 & (-13.0, -5.0)              & 41.70  (0.01)  &  0.71  (0.01)  &     3.58    (0.22)    & 0.118   &  13     &   120   &    0.1  &    0.3  \\
 P2 & ( -3.0,  4.0)$^{\rm 2}$    & 41.40  (0.02)  &  0.98  (0.05)  &     2.76    (0.23)    & 0.067   &  --     &   163   &    --  &    2    \\
    &                            & 42.60  (0.02)  &  0.74  (0.03)  &     3.82    (0.32)    & 0.067   &  --     &   152   &    --  &    2    \\
 P3 & ( -5.0,  7.0)$^{\rm 2}$    & 41.70  (0.04)  &  1.56  (0.08)  &     3.56    (0.29)    & 0.065   &  --     &   114   &    --  &    1    \\
    &                            & 42.70  (0.01)  &  0.68  (0.02)  &     3.59    (0.33)    & 0.065   &  --     &   106   &    --  &    1    \\
 P4 & ( -8.0,  5.0)              & 42.00  (0.01)  &  1.22  (0.02)  &     3.98    (0.26)    & 0.043   &  13     &   46    &    0.3  &    2    \\
 P5 & (  7.0,  4.0)              & 42.00  (0.02)  &  1.37  (0.03)  &     2.56    (0.33)    & 0.056   &  22     &   184   &    0.1  &    4    \\
 P6 & ( 13.0, -1.0)              & 42.70  (0.03)  &  1.69  (0.06)  &     2.89    (0.41)    & 0.049   &  29     &   148   &    0.2  &    4    \\
 P7 & (  8.0,  0.0)              & 42.40  (0.02)  &  1.33  (0.04)  &     3.14    (0.35)    & 0.058   &  21     &   252   &    0.1  &    4    \\
 P8 & (-21.0,  1.0)$^{\rm 3}$    & 42.40  (0.01)  &  0.5   (0.03)  &     5.56    (1.01)    & --      &   --    &    $<22$   & --    & --   \\
 P9 & ( 11.0, -6.0)              & 43.70  (0.02)  &  0.9   (0.04)  &     1.6     (0.53)    & 0.036   &  6      &   54    &    0.1  &    4    \\
P10 & (-15.0, 10.0)$^{\rm 4}$    & 42.20  (0.27)  &  1.25  (0.91)  &     0.98    (0.1 )    & --      &  --     &  $<16$  &    --     &    --   \\
P11 & (-15.0, 13.0)              & 42.00  (0.02)  &  0.6   (0.04)  &     2.97    (0.78)    & 0.023   &  2      &   19    &    0.1  &    5    \\
P12 & (  6.0, -1.0)$^{\rm 4}$    & 42.50  (0.02)  &  1.01  (0.03)  &     3.32    (0.43)    & --      &  --    &   95    &    --   &    --   \\
P13 & (-20.0, 17.0)$^{\rm 3}$    & 42.30  (0.02)  &  0.54  (0.05)  &     1.83    (1.18)    & --      & --    &    $<9$    &    --   &    --   \\
\hline
\end{tabular}
\end{center}
$^{\rm a}$ FWHM  from the hyperfine fits over an effective area of
radius R$_{\rm eff}$ deduced from CLUMPFIND \\
$^{\rm b}$ Total optical depth from the hyperfine fits  over an
effective area of radius R$_{\rm eff}$ deduced from CLUMPFIND  \\
$^{\rm c}$ Virial Mass, and 3.5~mm dust continuum mass computed over an
effective area of radius R$_{\rm eff}$ (deconvolved with the beam)
deduced from CLUMPFIND. Dust mass calculated for a temperature of 16~K.\\
$^{\rm d}$ virial parameter defined in the text\\
$^{\rm e}$ average gas density over  R$_{\rm eff}$\\
$^{\rm 1}$ \DAMM\ emission with associated very weak or no dust emission ($\le 3\sigma$). \\
$^{\rm 2}$ Since there are two velocity components, no virial mass has been calculated. \\
$^{\rm 3}$ point like \DAMM\ emission with associated very weak or no dust emission ($\le 3\sigma$). Dust continuum mass is only a 3$\sigma$ upper limit over the area integrated flux.\\
$^{\rm 4}$ point like \DAMM\ emission with associated continuum emission ($> 3\sigma$). \\
%Note: Similar notations  used  elsewhere in this paper.

\end{table*}

%TABLE 6.  NH2D/NH3 Derived parameters for PdBI data
\begin{sidewaystable*}[h]
\begin{center}
  \caption{ \AMM\ rotational temperatures, \DAMM\ and \AMM\ column
    densities, and fractionation toward G29.96e and G35.20w. '-'
    indicates non-detection at a 3$\sigma$ peak brightness in either
    tracer. For \DAMM\ detection with no \AMM\, we have used an
    excitation temperature of 16~K ($\pm 5$~K) for computing the
    column density. Ratios have been quoted only when the error
    estimate is below 50\%. \label{tab:deutn_ratio}}
\vspace{1em}

\begin{tabular}{clcccccccccccc}
\hline
Core$^a$ & Offsets  & \AMM\ & \DAMM\ & $v^{\rm NH_2D}_{\rm LSR}$ & $\Delta$v$^{\rm NH_2D}$ & $\tau_{\rm tot}$$^{\rm NH_2D}$ & $v^{\rm NH_3}_{\rm LSR}$ & $\Delta$v$^{\rm NH_3}$ & $\tau_{\rm tot}$$^{\rm NH_3}$& $T_{\rm rot}$   & $N_{\rm NH_2D}$        & $N_{\rm NH_3}$         & [\DAMM/\AMM]             \\
         & (arcsec) &       &        &  \kms\ &  \kms\ &  &   \kms\ &  \kms\ &  &(K)           & $10^{14} ~\rm cm^{-2}$ & $10^{15} (10^{14})~\rm cm^{-2}$  &                          \\
\hline
\multicolumn{5}{l}{\bf G29.96e \DAMM\ cores}\\
 P1 &  (  6.0, 25.0)&+ & + & 103.67 (0.01) &	0.87	(0.02) &	2.34	(0.29) &	102.4	(0.07) &	1	(0.13) &	0.82	(0.99) &         15.0 (8.1) &  3.5(2.0)   & 0.9	(1.1) & 	0.37	(0.11)   \\    	
 P2 &  (  1.0,  7.0)&+ & + & 103.17 (0.01) &	1.35	(0.02) &	3.71	(0.22) &	101.9	(0.05) &	2.09	(0.11) &	2.76	(0.45) &         22.6 (5.9) & 14.3(5.1)   & 7.9	(2.1) & 	0.17	(0.02)   \\    	
 P3 &  ( 22.0,-16.0)&- & + & 102.77 (0.02) &	0.67	(0.04) &	3.33	(0.77) &        	       &        	       &         &         16   (5)   &  3.9(1.6)   &           &                          \\ 	 
 P4 &  ( -3.0,-33.0)&- & + & 102.97 (0.03) &	1.15	(0.05) &	5.21	(0.83) &        	       &        	       &         &        16   (5)   & 10.3(4.0)   &            &                          \\	     
 P5 &  ( 9.0,-19.0)&+ & +  & 101.57 (0.01) &	0.6	(0.03) &	2.87	(0.76) &	100.5	(0.08) &	1.44	(0.22) &	1.74	(0.73) &       13.4 (3.2) &  2.7(0.9)   & 2.8	(1.3) & 	0.09	(0.03)   \\    	
 P7 &  ( -3.0,-28.0)&+ & + & 103.17 (0.02) &	1.01	(0.04) &	6.2	(0.77) &	100.6	(0.08) &	1.02	(0.35) &	4.9	(2.02) &        12.8 (4.7) &  9.3(3.0)   & 5.6	(3.1) & 	0.16	(0.07)   \\    	
 P10 &  ( -7.0, -4.0)&+& + & 102.47 (0.01) &	0.77	(0.04) &	0.55	(0.46) &	101	(0.03) &	1.39	(0.08) &	2.83	(0.45) &         14.9 (1.9) &  0.7(0.6)   & 4.4	(0.7) & 	       	         \\    	
 P8 &  (  2.0,-17.0)&+ & + & 103.57 (0.03) &	0.9	(0.06) &	3.56	(0.86) &	101.1	(0.04) &	1.81	(0.08) &	4.47	(0.56) &        15.7 (2.4) &  5.8(1.7)   & 9.1	(1.2) & 	0.06	(0.02)   \\    	
\multicolumn{5}{l}{\bf G29.96e  \AMM\ cores}\\
  & (  0.0,  6.0)&+ & +   & 103.17	(0.01) &	1.29	(0.02) &	3.82	(0.2 ) &        101.8	(0.04) &	2.02	(0.11) &	2.67	(0.43) &      22.1 (5.2) & 13.6(4.4)   & 7.2	(1.7) & 	0.18	(0.02)   \\    	            
  & (  2.0,-15.0)&+ & +   & 103.57	(0.03) &	0.9	(0.06) &	3.56	(0.86) &	101.1	(0.04) &	1.81	(0.08) &	4.47	(0.56) &       15.7 (2.4) &  5.8(1.7)   & 9.1	(1.2) & 	0.06	(0.02)   \\    	 
  & (-15.0,  10.0)&+ & -  &              &  		       &        	       &        101.4	(0.05) &	1.48	(0.13) &	2.33	(0.56) &        19.5 (5.0) &             & 4.2	(1.2) & 		      \\    	 
  & (-16.0, 28.0)&+ & -   &              &  		       &        	       &         94.7	(0.12) &	2.51	(0.27) &	 &       16.6 (7.2) &             & 7.9	(2.7) & 		      \\    	 
  & (  5.0,-29.0)&+ & -   &              &  		       &        	       &        101.0	(0.09) &	1.65	(0.19) &	3.27	(1.24) &       18.2 (8.7) &             & 6.4	(2.7) & 		      \\    	 
\multicolumn{5}{l}{\bf G35.20w  \DAMM\ cores}\\
P1 &   (-13.0, -7.0)&+ & + & 41.70	(0.01) &	0.69	(0.01) &	3.59	(0.24) &	41.8	(0.03) &	1.05	(0.06) &	2.96	(0.5 ) &          20.3 (4.9) &   6.1(1.9)  & 3.9	(0.9) & 	0.15	(0.02)   \\    	
P2 &   ( -4.0,  6.0)&+ & + & 42.60	(0.02) &	0.74	(0.02) &	3.68	(0.26) &	41.9	(0.02) &	1.57	(0.05) &	2.55	(0.23) &          19.0 (2.1) &   6.1(0.9)  & 4.8	(0.5) & 	0.12	(0.01)   \\    	
P2 &   ( -4.0,  3.0)&+ & + & 41.40	(0.02) &	0.93	(0.05) &	2.4	(0.21) &	41.9	(0.02) &	1.59	(0.05) &	2.59	(0.24) &          19.6 (2.2) &   5.2(0.9)  & 5.1	(0.6) & 	0.10	(0.01)   \\    	
P7 &   (  8.0, 0.0)&+ & + &  42.60	(0.03) &	1.68	(0.06) &	3.17	(0.37) &	42.5	(0.03) &	1.84	(0.06) &	2.4	(0.25) &        20.5 (2.6) &  13.3(2.7)  & 5.6	(0.7) & 	0.22	(0.03)   \\    	
P10 &  (-15.0, 14.0)&- & + & 42.10	(0.02) &	0.83	(0.04) &	2.44	(0.61) &		       &         	       &        &         16   (5)   &   3.5(1.5)  &               &                          \\ 	   
P8 & (-22.0, -1.0)$^{\rm 1}$&- & + &   42.30	(0.03) &	0.96	(0.06) &	3.41	(0.87) &	42.2	(0.06) &	1.26	(0.14) &	2.59	(0.89) &       16   (5)   &   5.7(2.5)  &               &                          \\  	   
\multicolumn{5}{l}{\bf G35.20w  \AMM\ cores}\\
   & ( 34.0, 0.0)&+ & - &                      &		       &	42.00	(0.02) &	1.00	(0.01) &	1.59	(0.33) &         20.9 ( 3.8) &            & 2.0	(0.5) & 	                 \\    	 
   & ( 20.0,  2.0)&+ & + &  42.16	(0.05) &	1.28	(0.09) &	1.28	(0.83) &	42.00	(0.03) &	1.73	(0.06) &	1.91	(0.25) &          20.0 ( 2.4) &   3.9(2.6) & 4.1	(0.6) & 		         \\    	
   & (  4.0, 0.0)&+ & + &   42.50	(0.02) &	1.41	(0.05) &	3.41	(0.39) &	42.30	(0.03) &	1.89	(0.06) &	2.75	(0.27) &         20.9 ( 2.9) &  12.3(2.7) & 6.7	(0.9) & 	0.17	(0.02)   \\ 
\hline
\end{tabular}
\end{center}
$^a$ Core names from the original \DAMM\ data (see
Table~\ref{tab:nh2d_line_parameter}) have been assigned to the cores
in the smoothed \DAMM\ data. \\
$^{\rm 1}$ \AMM\ (2,2) transition is not detected.
\end{sidewaystable*}
%%%%%########## DISCUSSION########################

\section{Analysis and discussion \label{sec:analysis}}
\subsection{Active high-mass star Formation in G29.96e and G35.20w}
As shown in Sect.\ref{sec:res_structure}, only the brightest mm cores at the center of both regions are actively forming
stars (i.e., in the mm1 cores of both G29.96e and G35.20w). The present paper does
not attempt to characterize the forming stars comprehensively; this
shall be the subject of forthcoming publications. Here, we limit
ourselves to a presentation of evidence that these forming stars are
likely to be massive (i.e., $> 8 \, M_{\sun}$).

First, both star-forming sites are located in the brightest dust cores
found in the respective map. They are massive; at 1.3~mm
wavelength, with the highest resolution available, we derive masses
of 18 and $291 \, M_{\sun}$ for the mm1 cores of G35.20w and G29.96e,
respectively, within radii of 0.03 and $0.16 ~ \rm pc$. Cores with such  masses are not found in
solar-neighborhood clouds only forming low-mass stars \citep{kauffmann2010c}. These clouds are typically associated with smaller mass cores:  8 and $76 \, M_{\sun}$ for the aforementioned radii
(based on Taurus, Perseus, Ophiuchus, and the Pipe Nebula; \citealt{kauffmann2010b:mass_size2}). This suggests that the stars that are forming within the mm cores in G29.96e and G35.20w are potentially massive.

\begin{figure*}
\centering
\begin{tabular}{cc}
\includegraphics[width=0.3\linewidth,angle=-90]{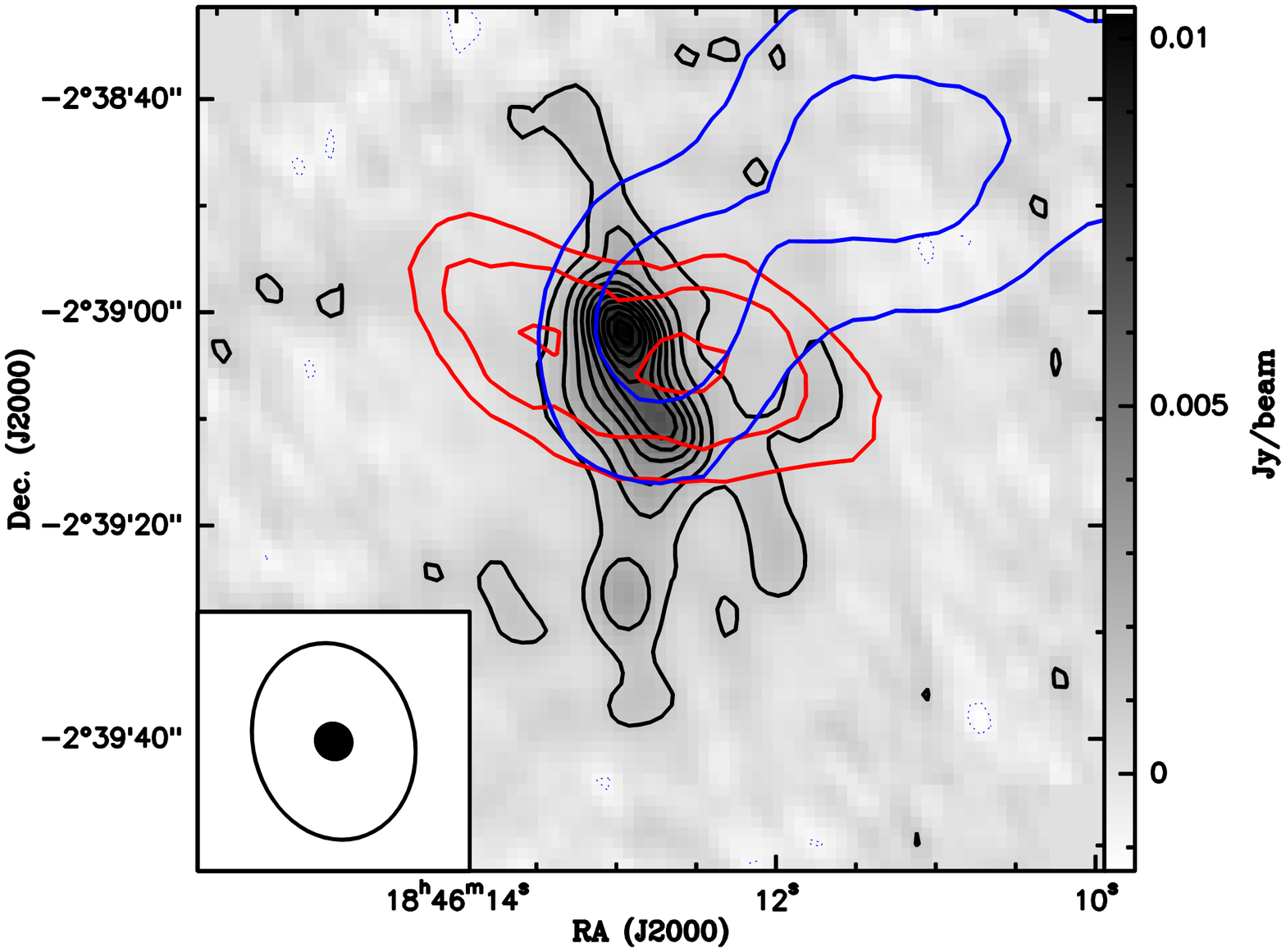} &
\includegraphics[width=0.3\linewidth,angle=-90]{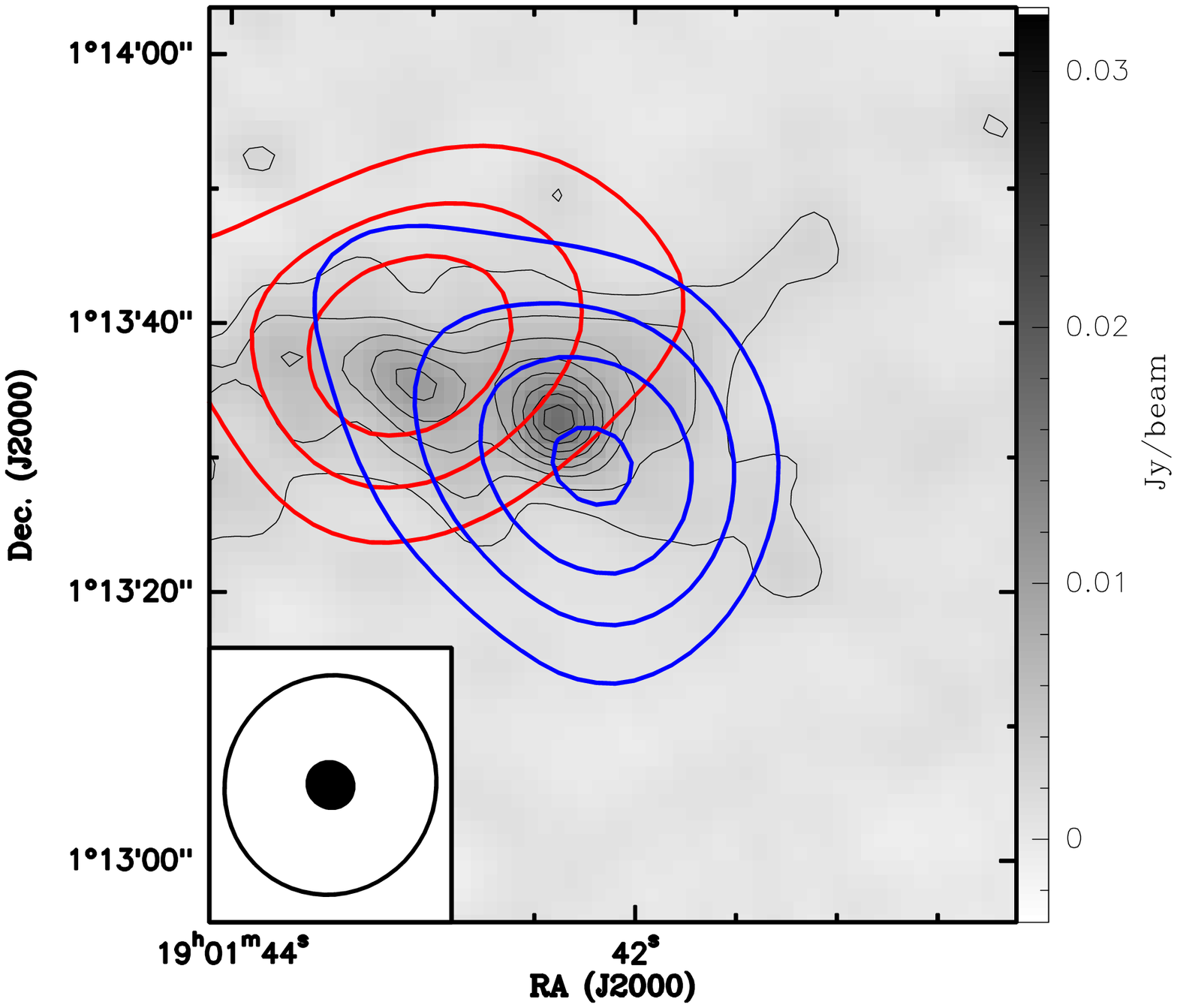}\\
\includegraphics[height=0.4\linewidth,angle=-90]{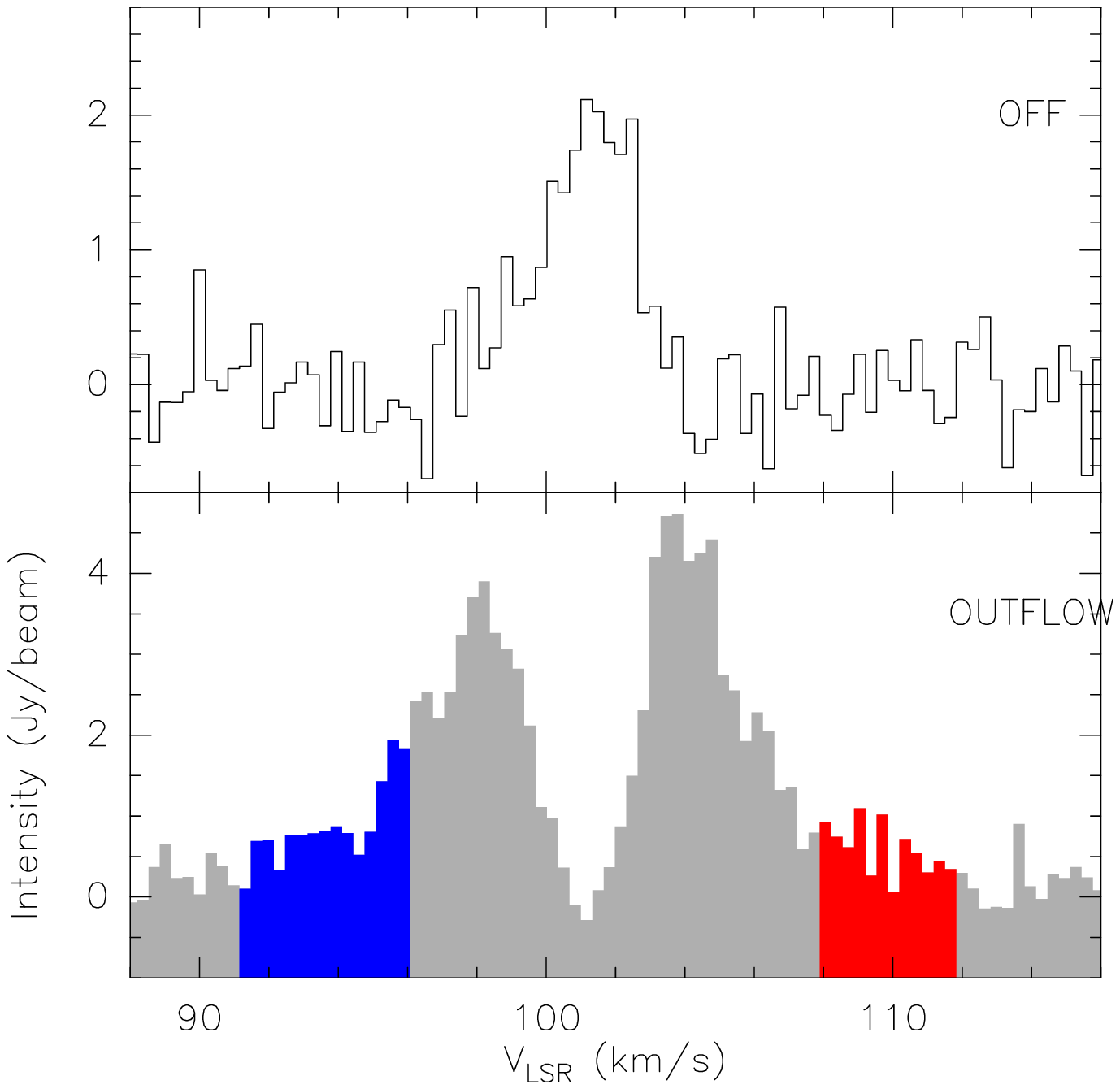} &
\includegraphics[height=0.4\linewidth,angle=-90]{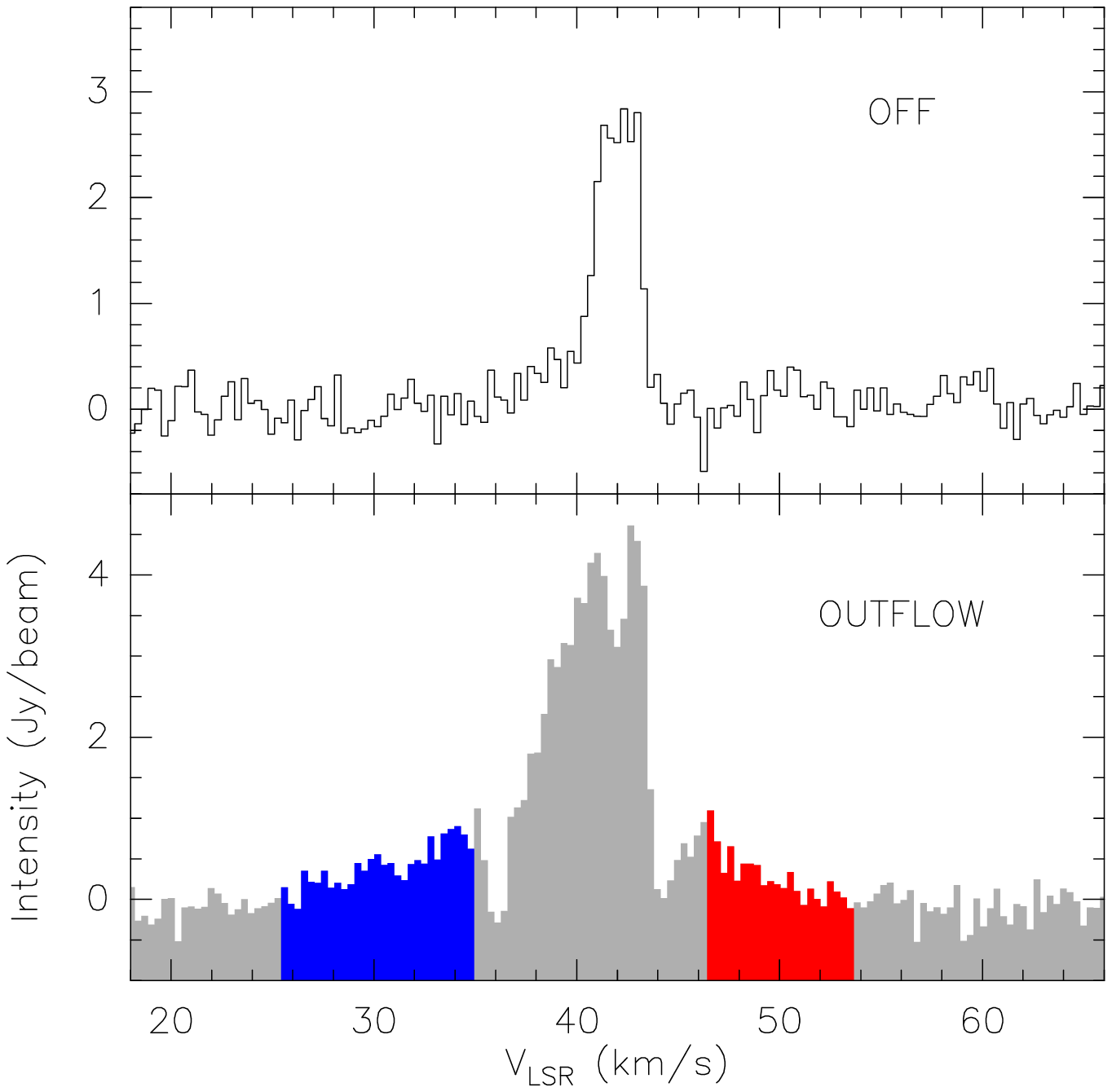}
\end{tabular}
\caption{{\it Top Left:} G29.96e PdBI 3.5~mm continuum plotted in gray-scales with \hco\ outflow overlaid in contours. Blue (red) contours represent the blueshifted (redshifted) shifted emission between velocities (LSR) of 91 -- 96~\kms (108 --
 112~\kms.) {\it Top Right:} G35.20w PdBI 3.5~mm continuum plotted in gray-scales with \hco\ outflow overlaid in contours. Blue (red) contours represent the blueshifted (redshifted) shifted emission between velocities (LSR) of 25.5 -- 35.0~\kms (46.6 --
 53.6~\kms.) {\it Bottom Panels:} The \hco\ spectra towards the brightest mm position mm1 (i.e. outflow source) and an OFF position for the respective sources in the top panel. The BIMA and PdBI beams are shown in white and black respectively.}
\label{fig:outflow}
\end{figure*}

Second, we detect outflows in both targets. As these outflows are detected at a
distance $>2$~kpc, the outflows must be massive, likely only from high
mass protostars.  In Figure \ref{fig:outflow}, we show the blue and redshifted \hco\ emission 
from our BIMA observations. This tracer is clearly associated with the
brightest mm cores in both objects (mm1). We find that there is
significant \hco\ emission from the cores and for clarity we show only the high-velocity outflow contribution, rather than include the bright systemic velocity emission from the cores. 
%This, together with
%the lowe(r) angular resolution of the data makes it difficult to
%properly image the \hco\ emission. \hco\ being the outflow tracer in
%our data set, we are interested in only deriving the outflow properties of mm1 in both sources. Therefore, we show only the outflow contribution. 
There is a clear, collimated redshifted and blueshifted component in both objects. 
Excluding the strong contribution from the core (or larger envelope)
from the outflow is tricky. To do this, we compared the outflow
spectra (around mm1) to that offset from mm1. Since outflows have
extended line wings, the velocity ranges that have outflow
contribution have been chosen by including only the broad line wings
on either sides of the LSR velocity of the cloud (constrained by our
\DAMM\ observations). The line wings extend to $\sim -10 (+11)$~\kms
for G29.96e, and $\sim -17 (+11)$~\kms  for G35.20w relative to the
systemic velocity of the gas. The velocity range for the blueshifted
(redshifted) emission is $\sim 5 (4)$~\kms for G29.96e, and $\sim 10
(7)$~\kms for G35.20w. We then derive the total gas mass contained in
the outflow. The \hco\ abundance is expected to be enhanced in
outflows \citep{rawlings04:hco+}. \citealt{zinchenko09:molecules} find
almost a constant \hco\ abundance in their sample of four high-mass
star-forming regions. We derive an average of 5.5$\times 10^{-9}$ from
their estimates, and adopt this value to calculate the outflow
mass. We calculate the \hco\ column density following \citet{caselli2002b} for an excitation temperature of 40~K. The total outflow mass is then 5.1~\Msol\ for G35.20w and 16.1 \Msol\ for G29.96e. A lower excitation temperature, e.g. 20~K, would lower the mass by $\sim 40\%$, however the outflow would still be characterized as massive. Such massive outflows are typical of high-mass protostellar objects \citep{beuther2002:outflows}. The mm1 sources in these two dark clouds are therefore harboring  even younger deeply embedded massive protostars or clusters.

 The same argument holds for Spitzer data, where we detect protostars at 24~$\mu$m towards the center cores in
both filaments (see Figure \ref{fig:nh2dspec}). Similarly, the detection of relatively bright $24 ~ \rm \mu m$ sources in
the target clouds suggests the presence of massive luminous stars. In G29.96e,
an aperture of $10 \arcsec$ radius to measure a flux density of $0.13 ~ \rm
Jy$; we subtract backgrounds, which are measured in an annulus inner and
outer radii of $10 \arcsec$ to $13 \arcsec$, respectively. For G35.20w, radii
of $13 \arcsec$ and $20 \arcsec$ are used to derive $2.0 ~ \rm Jy$. In both
cases, an aperture correction of 1.167 is used to boost the flux
densities\footnote{See MIPS Data Handbook Version 3 for details.}.
Given extended backgrounds, the measurements suffer from significant
uncertainties. These observations can be compared to the
luminosity-dependent $24 ~ \rm \mu m$ model flux densities of
\citet{dunham2008}, see \citet{parsons09} for details. We assume a $24 ~
\rm \mu m$ foreground extinction corresponding to $10^{23} ~ \rm cm^{-2}$,
as suggested by the peak column densities of our target clouds; much lower
luminosities apply otherwise. In this framework, the observed flux densities
imply luminosities of $6.8 \times 10^3 \, L_{\sun}$ (G29.96e) and $1.5 \times
10^4 \, L_{\sun}$ (G35.20w). This is in the range of luminosities expected for
massive young accreting stars \citep{sridharan2002:hmpo_catalog}.

Finally, Class~II methanol
masers at 6.7~GHz are detected towards one of the targets (i.e., G29.96e; Sect.\ref{subsec:res_g29structure}). These masers are exclusive signposts of the earliest stages
of high-mass star formation \citep{pestalozzi08:stat}.

\subsection{Line widths \label{subsec:linewidth}}
A fit to the resolved hyperfine structure of \DAMM\ allows for an
accurate estimate of the line width corrected for any optical depth
effects.  The observed line width is the sum of the thermal and
turbulent components, i.e. FWHM width $\Delta v$, is given by $\Delta
v^2 = \Delta v_{\rm th}^2 + \Delta v_{\rm nt}^2$. For gas temperatures
$< 20$~K, $\Delta v_{\rm th}<0.22$~\kms. The observed line width,
$\Delta v$, is in the range of 0.68 -- 1.42~\kms for G29.96e, and 0.5
-- 1.69~\kms for G35.20w. The mean and median line width representative of this sample is $\lesssim 1 ~ \rm km \, s^{-1}$. Therefore, the line width is only mildly supersonic
($\Delta v_{\rm nt}/\Delta v_{\rm th} \sim 4$) unlike usually those
expected from high-mass star-forming regions \citet{mckee02:100000yrs}. Recent high resolution observations by 
  \citet{olmi2010:vla_nh3} have also reported such
  low line widths in massive dense cores. This line width is also
smaller than the single dish line width measurements reported towards
high-mass IRDCs (for e.g. \citealt{pillai2006b:nh3},
\citealt{pillai2007}). Therefore, we are able (to some extent) resolve
the velocity dispersion within the beam, and we can ascertain that the
high-mass precluster cores are not necessarily characterized by large
velocity dispersions ($>1$~\kms). The implication of this property on
the core stability and current theories on high-mass star formation
will be discussed in Sect.~\ref{sec:theory}.

\subsection{Deuteration: NH$_2$D vs.\ NH$_3$ and
  dust\label{analysis:deutn}}
The importance of deuteration in our target clouds can be assessed by
deriving relative \DAMM{}-to-\AMM\ abundances,
$[{\rm NH_2D} / {\rm NH_3}]$. These are given in Table~\ref{tab:deutn_ratio}, calculated
using the scheme described in Sect.~\ref{sec:method_line_anal}. Towards the positions with a
signal strong enough to permit a reliable abundance measurement, we
derive $[{\rm NH_2D} / {\rm NH_3}] > 6\%$. As an average, we find
$\langle [{\rm NH_2D} / {\rm NH_3}] \rangle = 15\%$. The maximum
relative abundance is $37\% \pm 11\%$. This is high compared to the values reported in literature, for e.g. $[{\rm N_2D^+} / {\rm N_2H^+}]$, observed in low mass and intermediate mass prestellar cores (\citealt{crapsi2005:deutn_depn},  \citealt{fontani2008}) as well as $[{\rm NH_2D} / {\rm NH_3}]$ ratios observed in low mass  cores (\citealt{crapsi2007:nh2d}). More recently, \citealt{busquet2010:nh2d} have reported high deuteration ratios (0.1--0.8) in cores with no associated young stellar objects in a high-mass star-forming region.

Beyond the value of this abundance ratio, it is interesting to see how
the abundance of \DAMM\ varies across the clouds studied here.
Because of its particular formation chemistry, \DAMM\ can effectively
be used as a tracer of certain physical conditions. The remainder of
this section is dedicated to such studies.

\subsubsection{\DAMM\  chemistry  \label{analysis:deutn-theory}}
Let us first examine how deuterated Ammonia --- i.e.\ \DAMM\ --- is
supposedly produced. Two aspects warrant particular attention. First,
as laid out in \citet{pillai2007}, the deuterium fractionation
initiates  with the formation of \HTD\ through a reaction that plays a
role only at low temperatures ($\lesssim 20 ~ \rm K$). The enhanced
\HTD\ fractionation is then transferred to other molecules including \AMM.  Second, CO
depletion by freeze-out onto dust grains \citep{tafalla2002:depletion} aids the
formation of \DAMM{}, since gas-phase CO ``destroys'' \HTD\, and
therefore \DAMM\ \citep{caselli2008:h2d+}.  In addition, \AMM\ does
not deplete considerably relative to other molecules. Given similar
volatility of the N$_2$ (parent molecule for \AMM\ and \DAMM) and CO,
it is yet unclear why. The deuterium fractionation however is
sensitive to temperature such that at temperatures higher than 20 ~K
(close to protostars),
release of CO to gas phase together with lower fractionation would
lead to a decrease in the \DAMM\ deuteration.
In summary, \DAMM\ is thus expected to be abundant in cold regions with significant CO
depletion.

\subsubsection{Spatial abundance patterns \label{subsec:nh2d_vs_dust}}
\begin{figure*} \centering
\includegraphics[width=0.8\linewidth,angle=0]{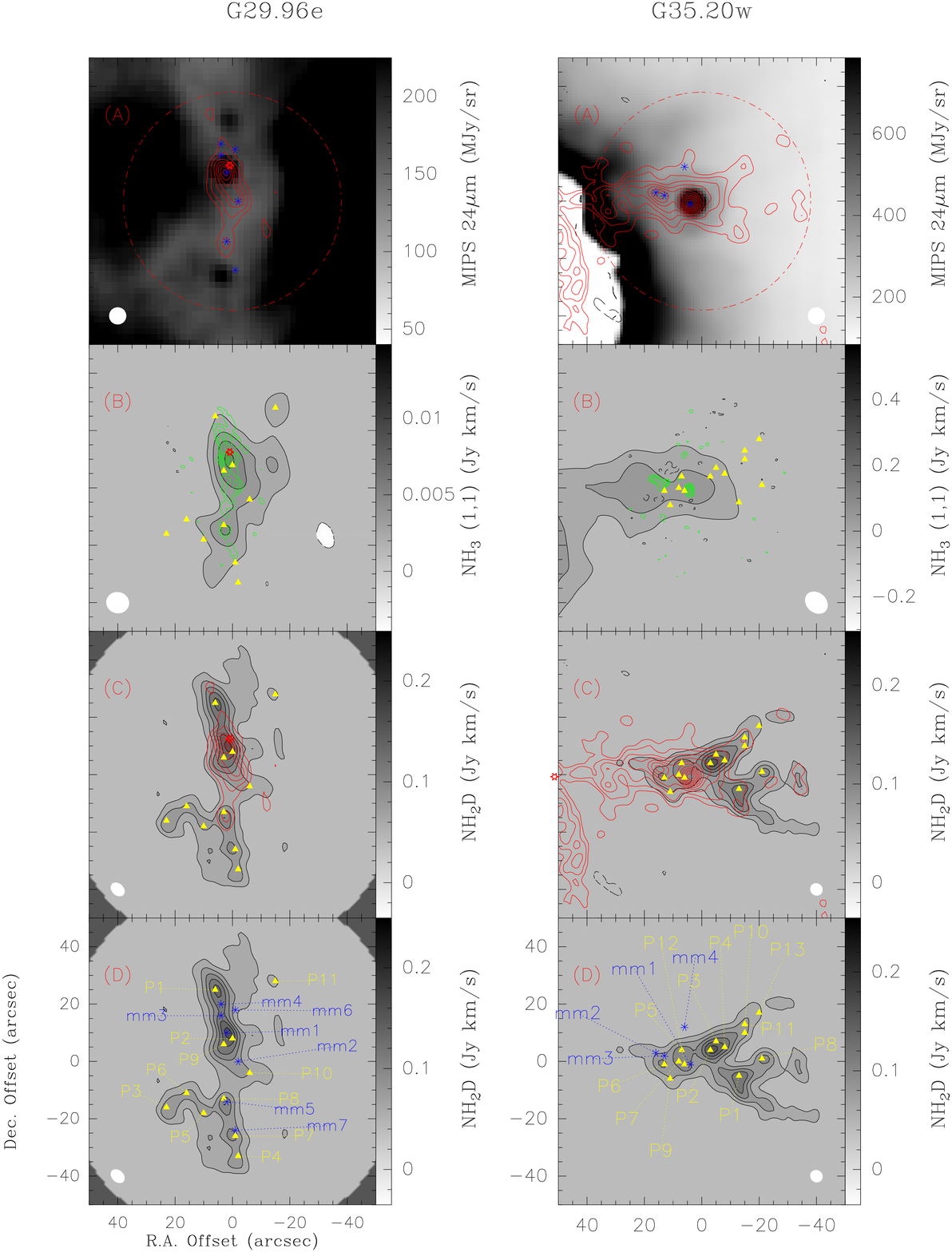}
\caption{Line and Continuum emission towards G29.96e (\textit{left}) and
  G35.20w (\textit{right}).  \textit{Panel (A)}: Spitzer 8$\mu$ (for
  G29.96e)/24$\mu$ (for G35.20w) emission and 3.5~mm PdBI dust
  continuum emission as contours. \textit{Panel (B)}: VLA \AMM\
  (1,1) integrated intensity map and  1.3~mm PdBI dust
  continuum emission as contours. \textit{Panel (C)}: PdBI \DAMM\
  integrated intensity map and 3.5~mm PdBI dust
  continuum emission as contours.  \textit{Panel (D)}: PdBI \DAMM\
  integrated intensity map with the clumps identified from CLUMPFIND
  from P1. The PdBI primary beam at 3.5~mm is shown as dashed circle in the top most panel. The contour levels
  for all the maps (line and continuum) start at $-3\sigma$, $3\sigma$
  in steps of 3$\sigma$.  The positions of the \DAMM\ cores (P1..) given in
  Table~\ref{tab:nh2d_line_parameter} are marked as filled
  triangles, the 1.3mm cores (mm1..) are marked as stars and the
  synthesized beam of the background image shown as filled white ellipses. For G29.96e, we have also marked the methanol maser
  positions identified by \citet{walsh1998:meth_radio} as red stars.}
\label{fig:g29_g35-all}
\end{figure*}

Dust emission and infrared extinction are good probes of $\rm H_2$
column density. The coldest cloud regions, and the regions with
strongest CO depletion,are usually strongly correlated with $\rm H_2$
column density. Therefore, a basic prediction of the formation
theory of deuterated analogues of dense gas tracers like \NTH\, and \AMM\ is that they should correlate well with dust emission
intensity and extinction features. As seen in Figs.\ref{fig:g29_g35-all}, this is, in
a global sense, the case in G29.96e and G35.20w. There are, however, some
interesting exceptions to this general trend.

When examining the correlation of emission from dust, \AMM{}, and \DAMM\ in
more detail we see in Fig.\ref{fig:overview} that the velocity-integrated \AMM\
emission correlates very well with the PdBI dust emission map at 3.5~mm
(the map at 1.3~mm wavelength is not useful for this because of spatial
filtering). This correlation holds in terms of spatial distribution
(i.e., dust and \AMM\ are present in the same regions), as well as in
terms of intensity (i.e., very bright regions in dust are also bright in \AMM{}, and vice versa). This situation is
entirely different for \DAMM{}; the dust emission peak is not the brightest \DAMM\ peak in both filaments, 
thus the brightest dust emission locations do
not dominate the \DAMM\ maps. To give an example, the
\DAMM{}-identified P1 position in G29.96e is free of PdBI 3.5~mm dust
emission, while P2 coincides with the mm1 dust emission peak; still,
both positions are similar in \DAMM\ brightness. Also, many major
\DAMM\ peak locations (e.g., P1, P2--P4, and P8 in G35.20w) are not
unusually bright in 3.5~mm dust emission; sometimes (e.g., P1 in G29.96e),
no dust emission at all is detected. Correspondingly (because dust and
\AMM\ have a similar distribution), the detailed correlation of \DAMM\
and \AMM\ is not perfect.

\begin{figure*}
\centering
\begin{tabular}{cc}
\includegraphics[width=0.4\linewidth,angle=-90]{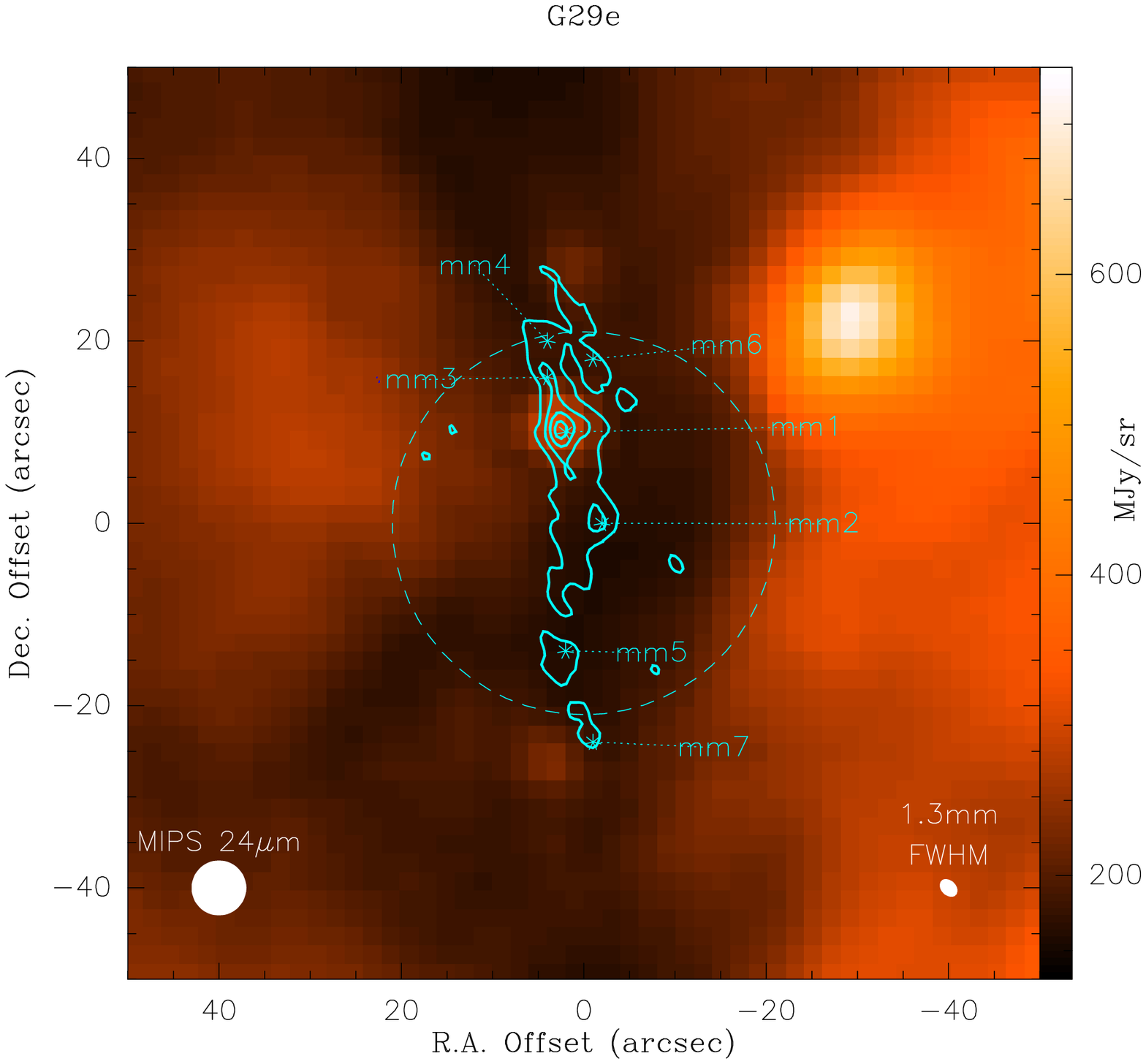} &
\includegraphics[width=0.4\linewidth,angle=-90]{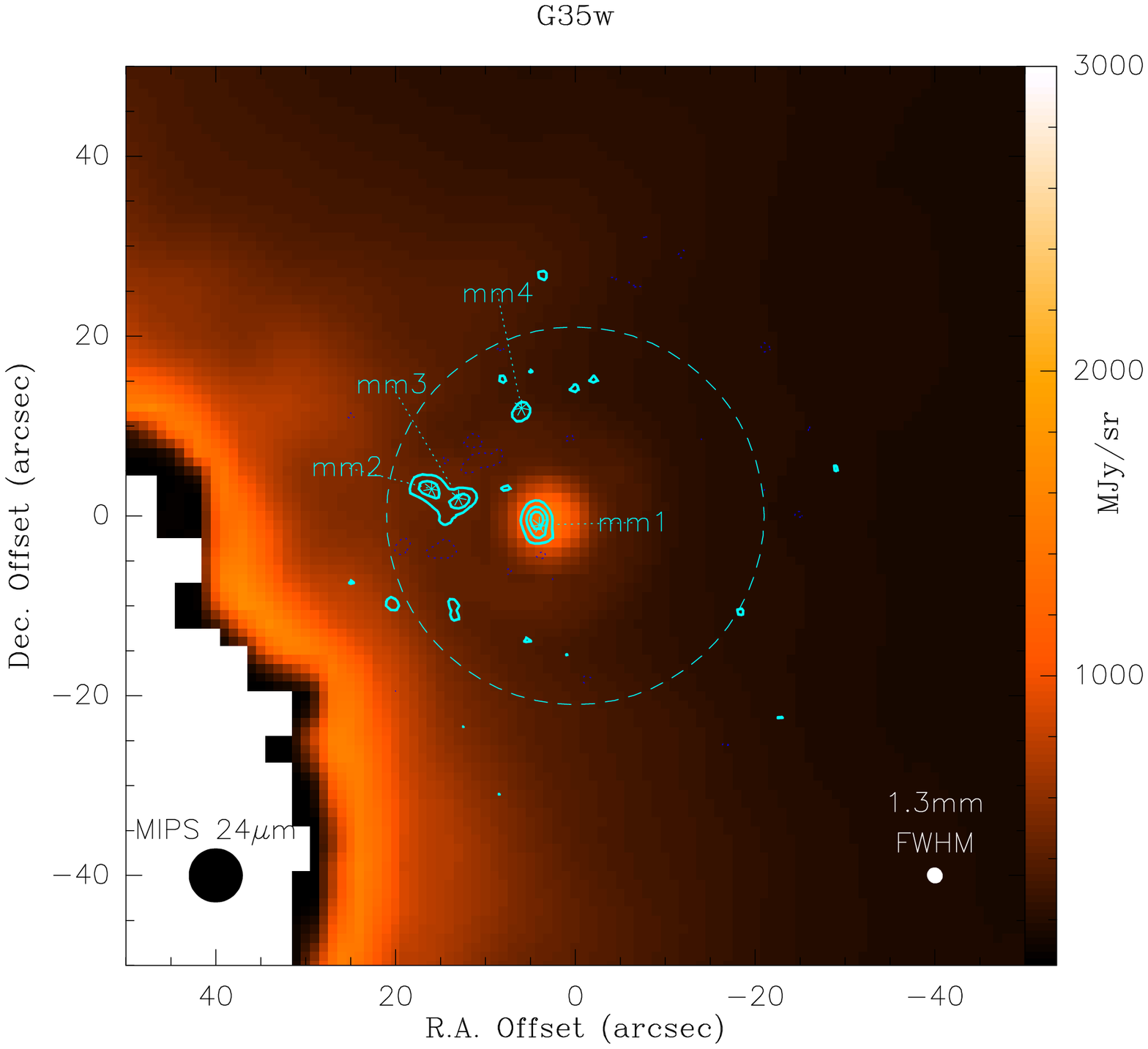}\\
\includegraphics[width=0.38\linewidth,angle=-90]{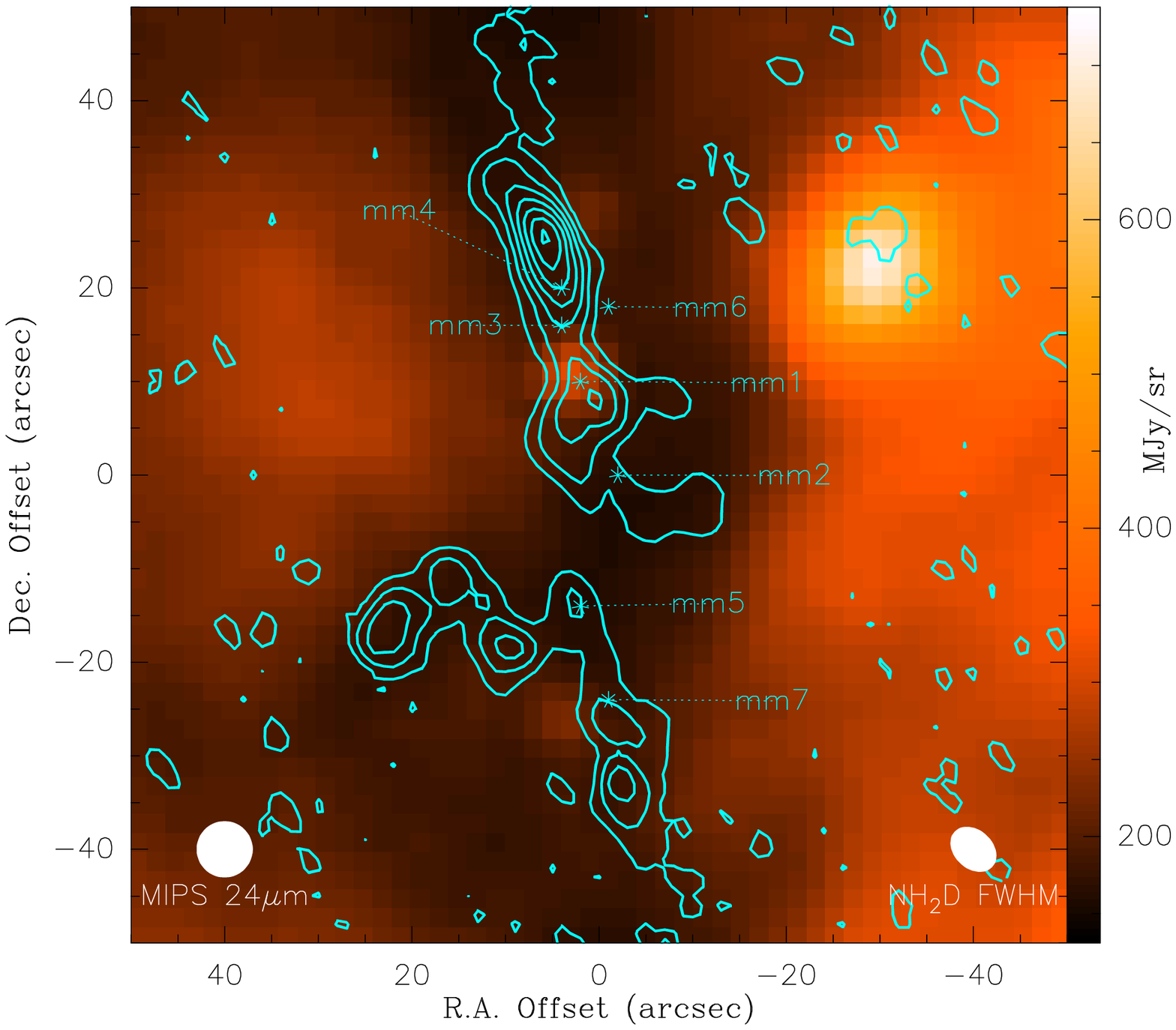} &
\includegraphics[width=0.38\linewidth,angle=-90]{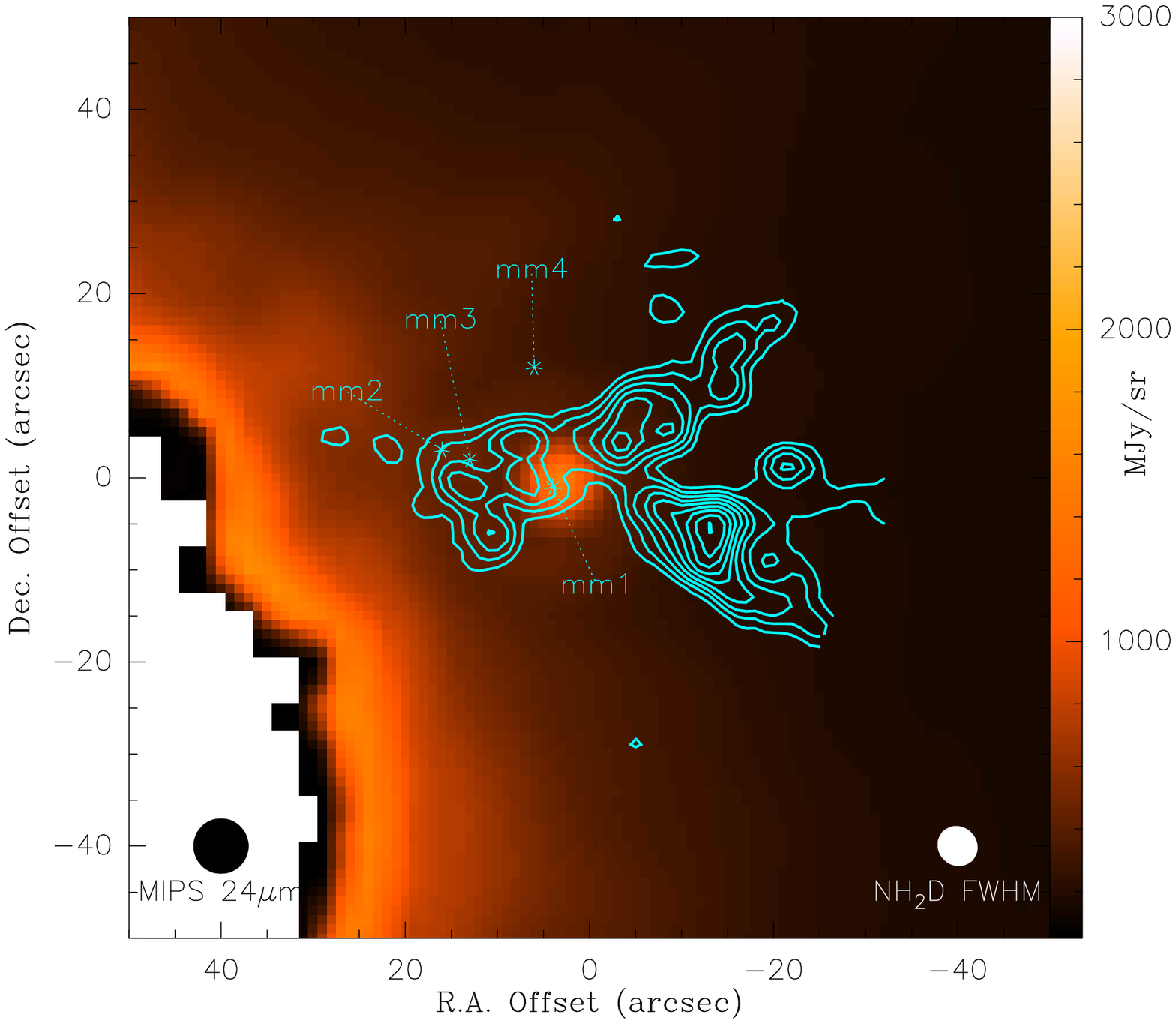}\\
\includegraphics[height=0.4\linewidth,angle=-90,bb=195 47 553 684,clip]{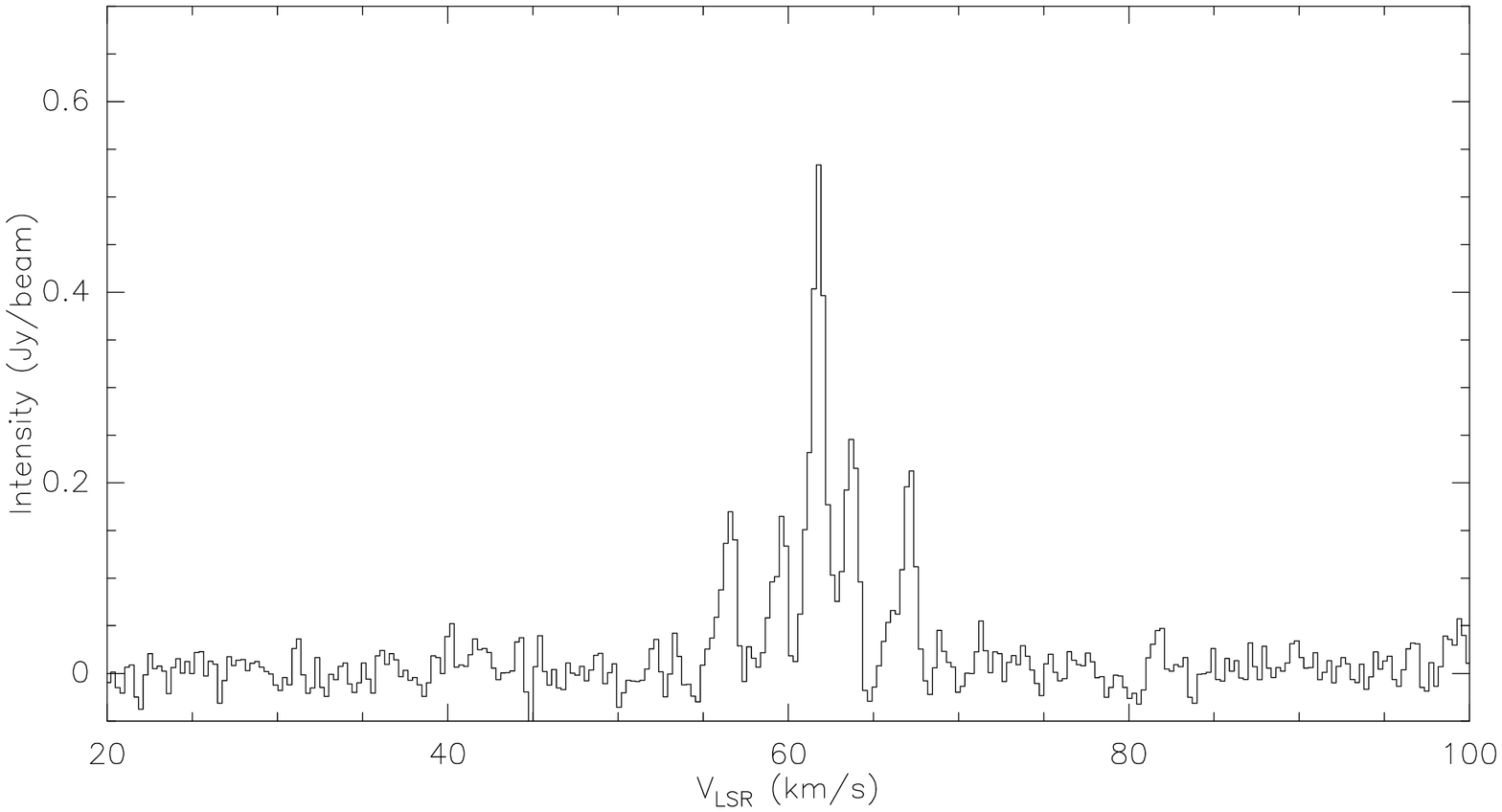} &
\includegraphics[height=0.4\linewidth,angle=-90,bb=195 47 553 684,clip]{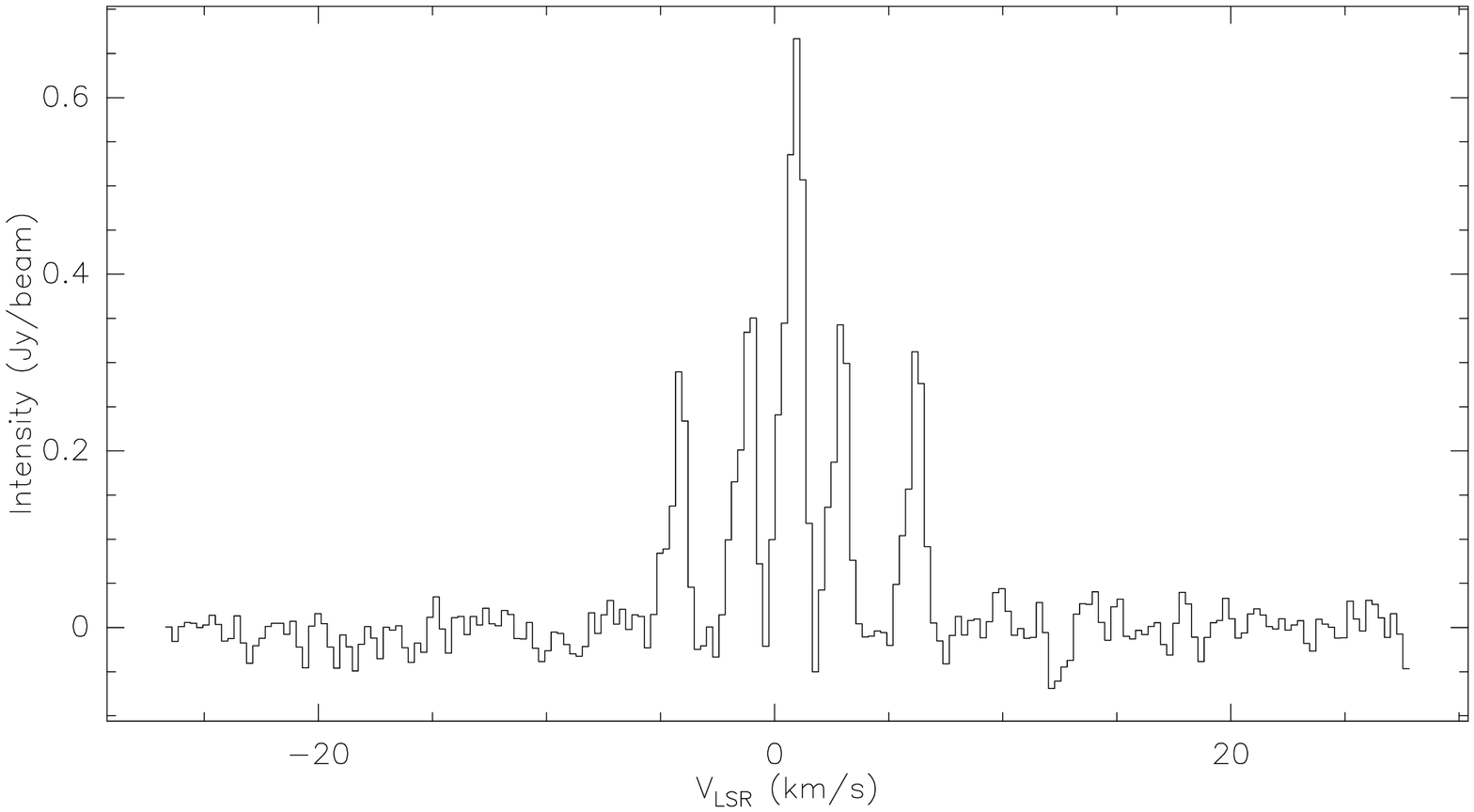}
\end{tabular}
\caption{ {\it Top Panels:} Spitzer 24$\mu$m map (color scale), and
  PdBI 1.3~mm emission intensity map with the PdBI 1.3~mm primary beam indicated as dashed circle. The contour levels start at $-3\sigma$, $3\sigma$ in steps of 5$\sigma$. {\it Center Panels:} Spitzer 24$\mu$m map (color scale), and
  PdBI \DAMM\ peak intensity map. The \DAMM\ contour levels start at $-3\sigma$, $3\sigma$ in steps of 2$\sigma$. 
 {\it Bottom Panels:} The \DAMM\ spectra towards the brightest
 position  for the respective sources in the top panel. The
 brightest \DAMM\ peak is is clearly offset from the Spitzer point
 source and  brightest mm core in both cases, and that the clumps are detected at high S/N through out the map.}
\label{fig:nh2dspec}
\end{figure*}

To understand these trends, first consider the dust emission peak
positions. These are heated by the embedded forming stars; \AMM\
observations (Table \ref{tab:nh2d_line_parameter}) yield gas temperatures $\approx 22 ~ \rm K$
for the mm1 positions in G29.96e and G35.20w. As discussed in
  Sect.\ \ref{analysis:deutn-theory}, at these temperatures, deuterium
fractionation is expected to drop. Also, the heating can lead to evaporation of CO from dust
grains, which will further reduce the \DAMM\ abundance.  The dust
emission peaks are thus actually not expected to be particularly rich
in \DAMM{}, in case they contain stars \citep{emprechtinger09}. The PdBI \DAMM\ peak 
intensity map overlaid on the Spitzer 24$\mu$m image in Fig.\ref{fig:nh2dspec} shows a clear offset in the peak position 
towards the protostar seen in emission at 24$\mu$m in both filaments.

Next, we consider the case of the relatively faint ($\lesssim 9\sigma$) \DAMM\ peak positions free of detected dust
emission. Towards those positions, dark extinction features in the 8
and $24 ~ \rm \mu m$ Spitzer images indicate significant dust column
densities. However, the dust emission sensitivities reported in
Table~\ref{tab:beam_rms} imply $1~\sigma$ column density detection
thresholds $\sim 5\times 10^{22}$~\cmsq\ for a temperature of
15~K. Thus, it is not surprising that the dust emission towards
relatively faint \DAMM\ cores goes undetected. In fact, a
  recent 450$\mu$m image of G35.20w taken as part of SCUBA-2
  commissioning clearly show the \DAMM\ streamers that go undetected with
  PdBI dust continuum \citep{jenness2010}. Similarly, For G29.96e, the
filamentary structure observed in \DAMM\ coincide with the dark
Spitzer extinction features.
In summary, the spatial trends observed in \DAMM\
emission are thus in broad agreement with the theoretical picture
drawn in Sect.\ \ref{analysis:deutn-theory}.

\subsection{Mass estimates\label{subsec:mass}}
 We derive the gas mass from the 1.3~mm and 3.5~mm dust
continuum emission as detailed in Sect.\ \ref{subsec:mass_method}).  In principle these masses, which were obtained from the flux
  integrated over a certain area might be influenced by foreground and
  background emission along the line of sight to the
  observer. Estimates for density gradients typical of clouds
  (Kauffmann et al., in prep.) suggest that the mass of embedded cores
  is usually overestimated by up to a factor 2. However, in our maps,
  any large scale information is lost in the 1.3~mm data because
  structures larger than $\approx 19$\arcsec\ are effectively filtered
  out by the interferometer. In practice, both biases cancel each
other to some extent, and the derived masses and densities are
likely to be correct within a factor $\sim{}2$ (when setting uncertainties due to opacities and temperatures aside).

We can quantify the extent of missing flux by comparing the total mass
within an aperture for SCUBA and PdBI continuum data.  For an aperture
with a diameter of 38~\arcsec, which corresponds to the greatest extent of
the N-S structure seen in SCUBA data for G29.96e, we estimate the total
mass within the filaments to be $2300 \, M_{\sun}$ (G29.96e), and
$1050 \, M_{\sun}$ (G35.20w) from the 3.5~mm data. The single-dish SCUBA continuum mass is derived to be $3550 \, M_{\sun}$ (G29.96e), and
$2250 \, M_{\sun}$ (G35.20w). 
Therefore, we conclude that the observations suffer
from 50 -- 65\% missing flux on the shortest baselines of the
interferometer at 3.5~mm. For G35.20w SCUBA measurements, we used archival SCUBA data because of significant artifacts in our data. 

We can also compare the masses derived from the 1.3mm and 3.5mm data.
In G29.96e, clumps of masses $\sim 200 ~ {\rm to} ~ 1000 \, M_{\sun}$ are
identified in the 3.5~mm dust continuum images. Our 1.3~mm observations,
however, reveal twice as many clumps as have been identified at
3.5~mm. This might be because G29.96e is at a distance of 7.4~kpc, so that
substructure is easily confused in the 3.5~mm beam. As shown in
Table~\ref{tab:pdbi_cont_mass}, the total mass in 1.3~mm clumps of G29.96e
exceeds the total mass in 3.5~mm clumps by a factor 1.3. In G35.20w,
however, this ratio is 0.1;g much lesser mass is detected at 1.3~mm,
probably because of filtering. This may not be surprising, given that
G35.20w is closer by a factor 2.\medskip

\noindent{}Virial masses, derived from the \DAMM\ data, provide an independent
mass estimate. Following the framework of
\citet{bertoldi1992:pr_conf_cores}, we define the virial mass as
\begin{equation}
M_{\rm vir} = \frac{5 \sigma^{2} R}{G} \, ,
\label{eq:viriall-mass}
\end{equation}
where $\sigma$ is the 1-dimensional velocity dispersion, $R$ is the
radius of the clump, and $G$ is the constant of gravity. The velocity
dispersion is calculated as
$\sigma^2 = \sigma_{\rm th}^2 + \sigma_{\rm nt}^2$, to include thermal
and non-thermal contributions to the supporting motions (Sect.\
\ref{sec:fragmentation}). Velocity dispersions, $\sigma$,
derive from FWHM widths, $\Delta v$, as
$\sigma = (8 \ln[2])^{-1/2} \, \Delta v$. We stress that
  $M_{\rm{}vir}$ is a characteristic property of a cloud, but is not a
  strict estimator of the actual mass. Indeed, virial and dust masses
can be compared using the virial parameter,
\begin{equation}
\alpha = \frac{M_{\rm vir}}{M} = \frac{{5}{\sigma^{2}}{R}}{G M} \, .
\label{eq:alpha}
\end{equation}
As we see in Table~\ref{tab:pdbi_cont_mass} (and discussed further below in Sect.\ \ref{subsec:vir}), $\alpha \ll 1$ for all cores which means that these cores are gravitationally over-bound. The virial masses do thus provide a lower limit to the estimated
dust masses. We provide the temperature, size (after deconvolution), column
density, volume density, mass, virial parameter, and aperture mass of the identified clumps
 in Table.~\ref{tab:pdbi_cont_mass}.

\subsection{Correlations between mass and temperature}
\begin{figure}
\centering
\includegraphics[height=5cm,angle=0]{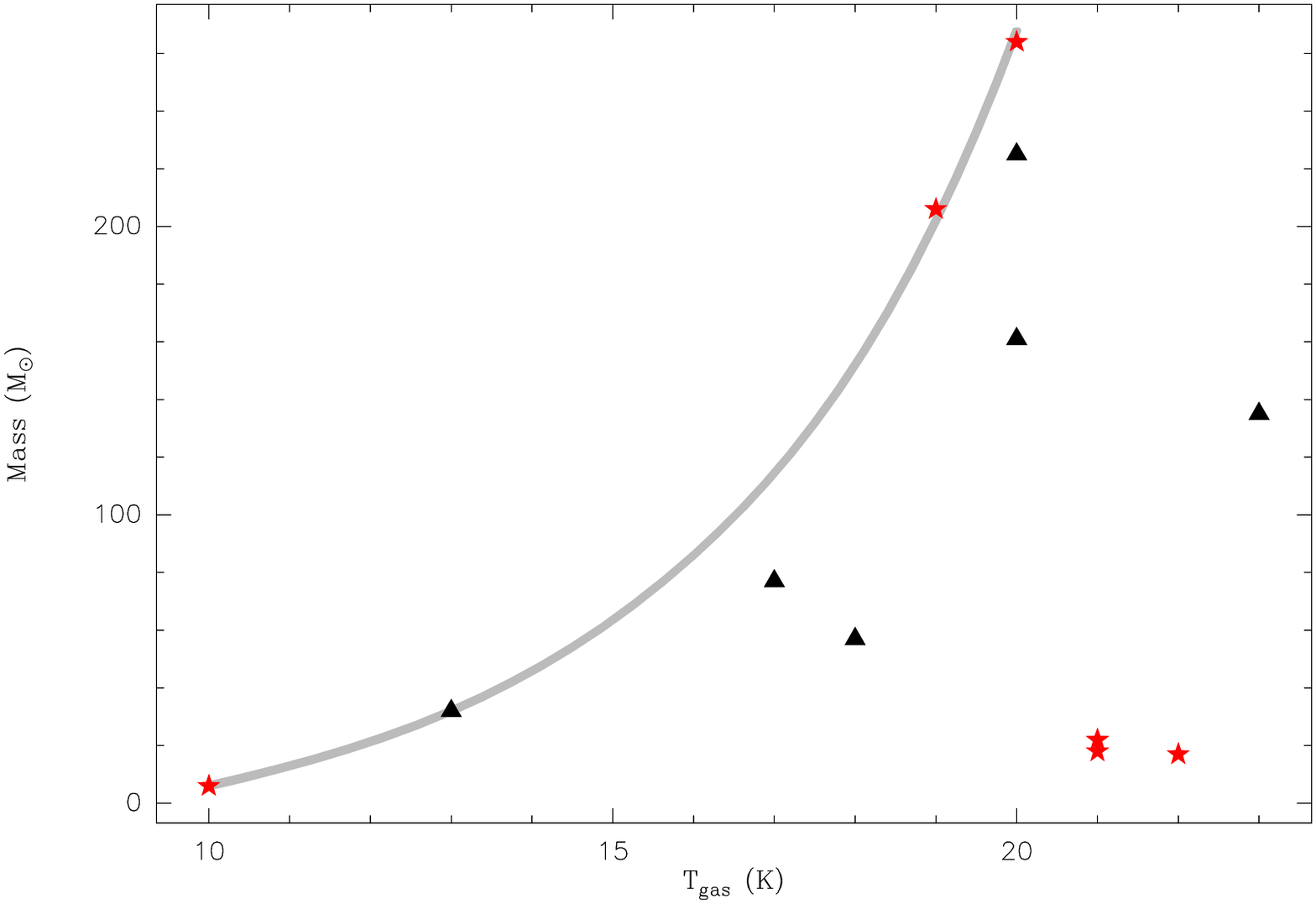}
\caption{Gas temperature derived from VLA \AMM\ observations versus
  the dust mass for the two filaments. Triangles and stars mark
  G29.96e and G35.20w respectively. The gray curve is shown to indicate the parameter space
  avoided in mass-temperature space.}
\label{fig:tkin-mass}
\end{figure}

As  discussed in Sect.\ref{subsec:nh2d_vs_dust}, the brightest dust
cores are not the brightest in \DAMM. Are these bright dust continuum
cores (see Table~\ref{tab:pdbi_cont_mass}) exceptional?
Figure \ref{fig:tkin-mass} explores the relation between the \AMM{}-derived gas
temperatures and the dense core masses (from PdBI dust continuum
observations) for these cores. To remove biases due to different source distances, the
masses are all calculated for an aperture of 37000~AU diameter. Here, 37000~AU corresponds to the PdBI 3.5~mm beam of $\sim$~5 arcsec for the more distant source G29.96e.

As we see, the maximum core mass that we derive increases with increasing gas
temperature; cores of low mass, for example (i.e.,
$\le 40 \, M_{\sun}$) are found in all temperature domains, but all
massive cores (i.e, $\ge 150 \, M_{\sun}$) have high temperatures
(i.e., $\ge 18 ~ \rm K$). Most likely, the heating needed to raise the
temperatures comes from heating by embedded stars. Heating to
temperatures $\gtrsim 10 \, K$ can be provided by either
(\textit{i}) a single star luminous (and thus massive) enough to
provide the heating, or (\textit{ii}) a cluster of low-mass stars
that is populous enough to have a high luminosity. The luminosity required for the heating is not high; following Eqs.\
(5, 8) of \citet{kauffmann2008}, $100 \, L_{\sun}$ are, e.g., enough
to raise the mean dust temperature within a 37000~AU diameter aperture
to 20~K.

Regardless of the nature of the sources might be, the dearth of cold
(e.g., $\le 15 ~ \rm K$) high-mass cores ($\ge 100 \, M_{\sun}$)
suggests that all cores of such high-mass might form stars. These are not
entirely devoid of stars and do not exactly represent the conditions
before the onset of active star formation. Rather, it is the bright \DAMM\ cores \emph{that are on
average colder} which represent the initial conditions of high-mass star formation.

\subsection{Fragmentation\label{sec:fragmentation}}
When trying to understand the initial conditions of (high-mass)
star-formation, cloud fragmentation is one of the most important
properties to be studied \citep{field08}.  Indeed, our interferometer
observations show that each SCUBA clump fragments into several cores.
The initial mass function, star formation efficiency, and star
formation rate all depend on cloud fragmentation. To this end, we looked at the fragmentation in our sample from the core separation and
their masses (Fig.\ref{fig:fragmn}).

\subsubsection{Theory}
\begin{figure}
\centering
\includegraphics[height=7.5cm,angle=-90,bb=86 32 508 707,clip]{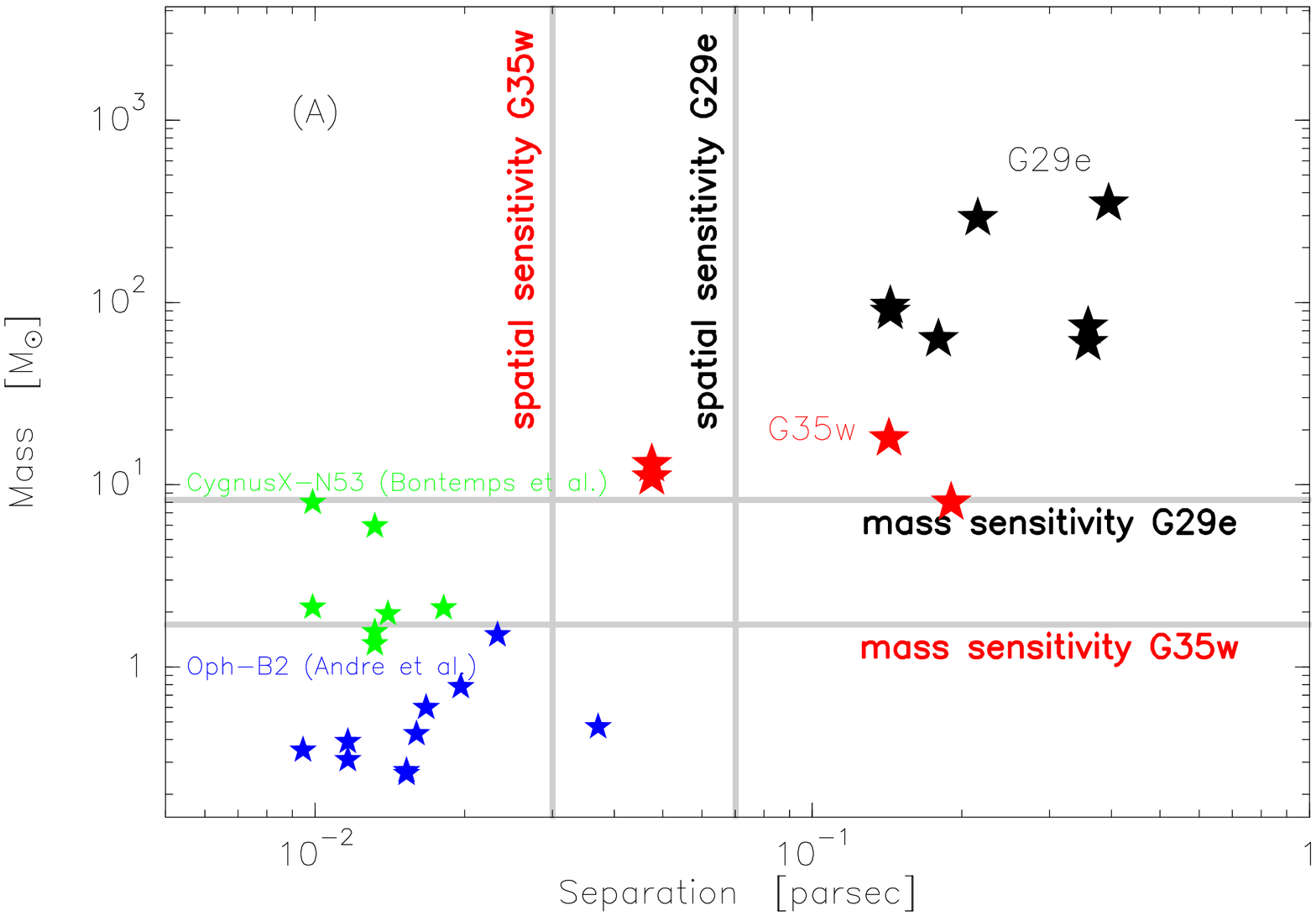}\\
\medskip
\medskip
\includegraphics[height=7.5cm,angle=-90,bb=86 32 508 707,clip]{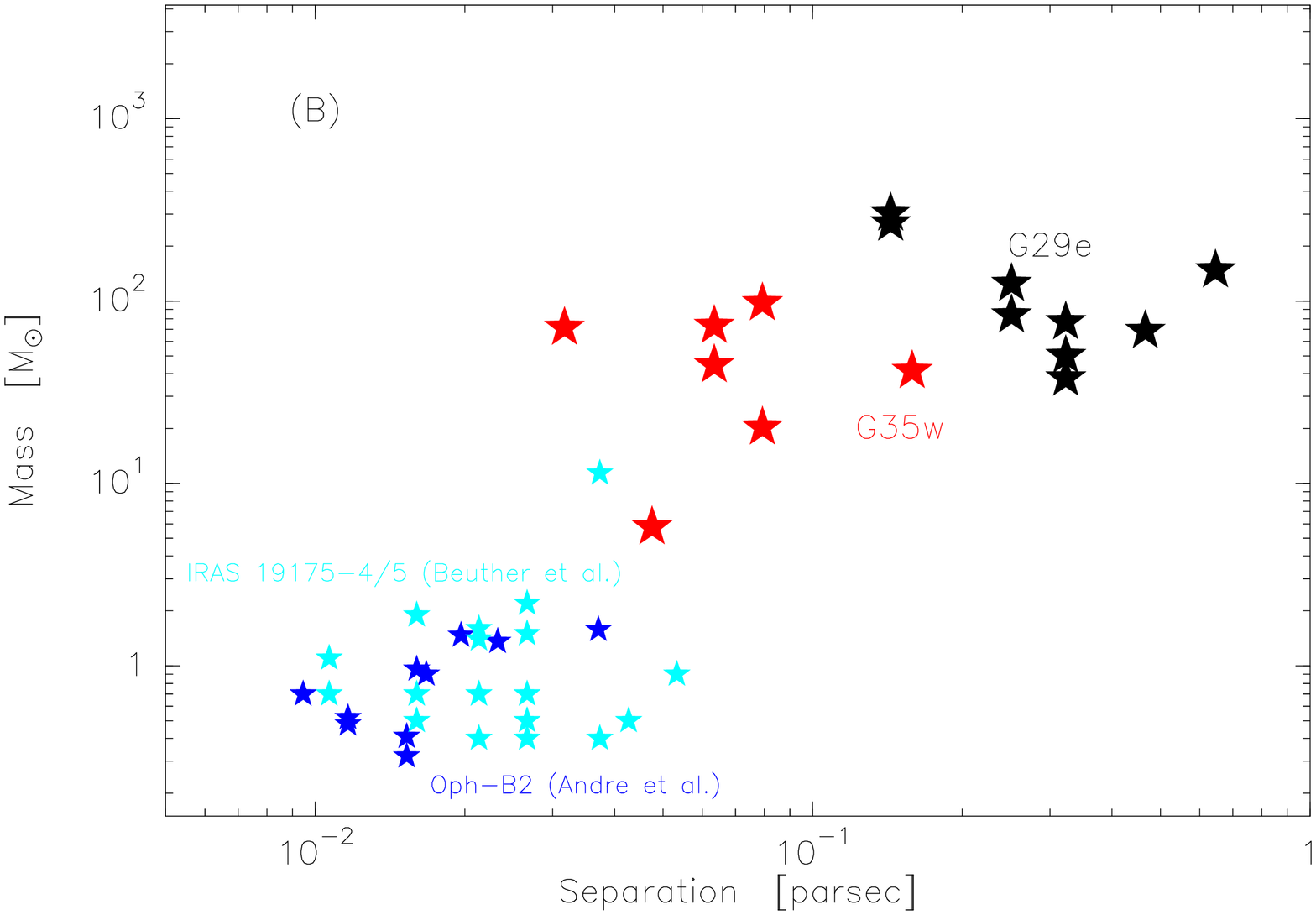}\\
\includegraphics[height=7.5cm,angle=-90,bb=86 32 508 707,clip]{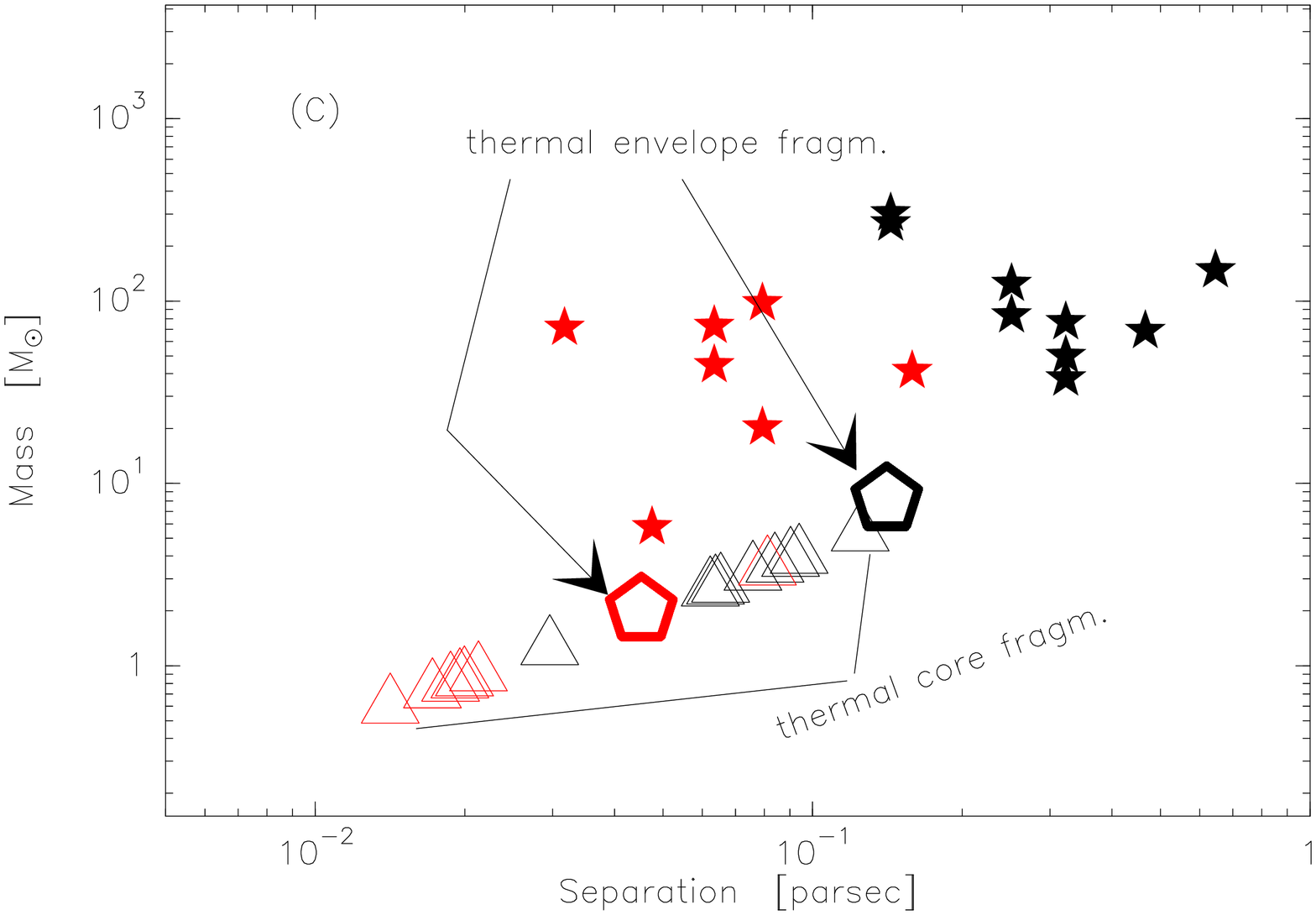}\\
\includegraphics[height=7.5cm,angle=-90,bb=86 32 555 707,clip]{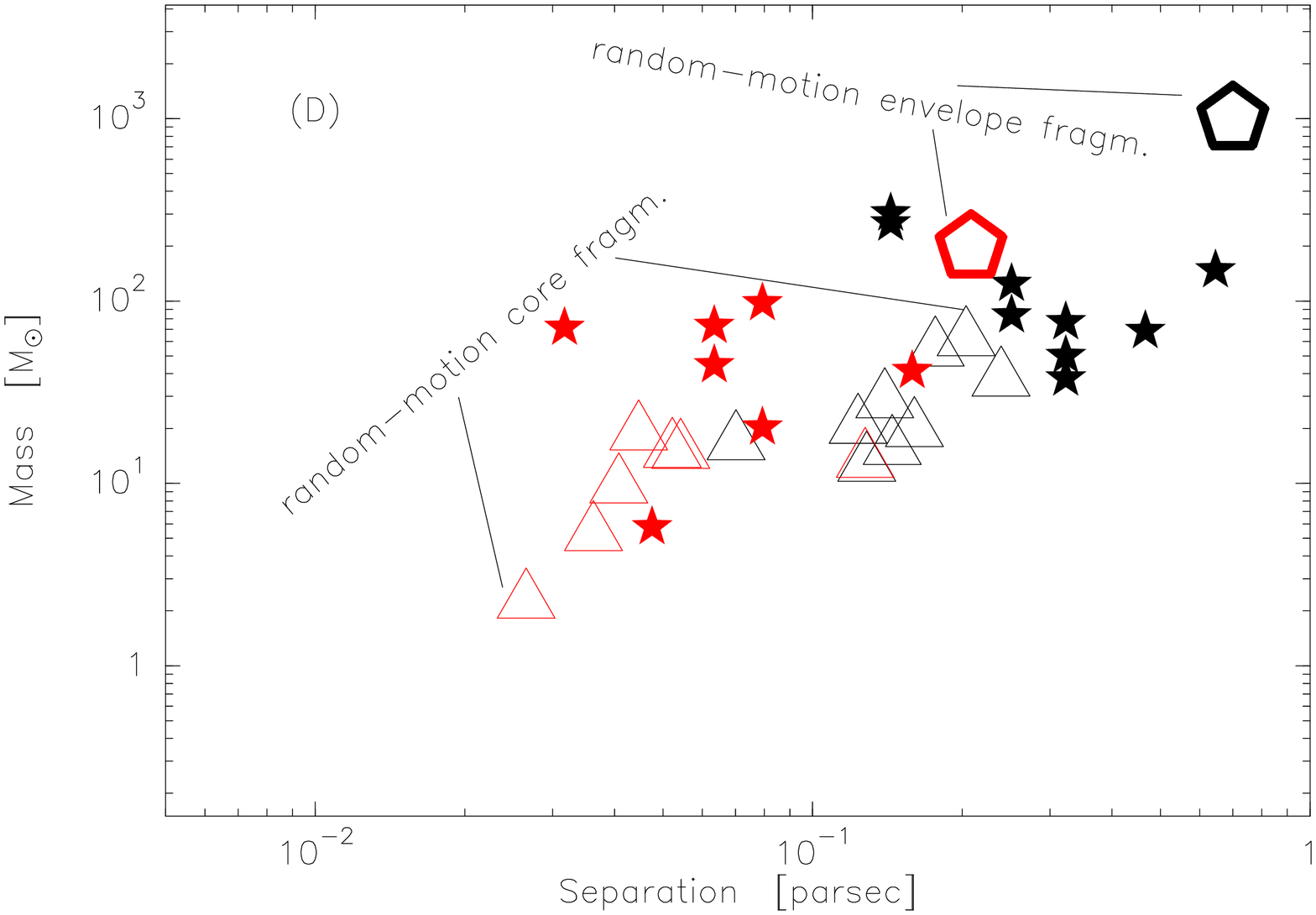}\\
\caption{Separation versus core mass. For all panels, stars mark the
  observations, pentagons mark the envelope properties for
  thermal/turbulent Jeans fragmentation while triangles mark the core
  properties. \textit{Panel (A)}: We used 1.3~mm PdBI
  data for G35.20w(red) and G29.96e(black). The mass
  sensitivity corresponds to a 3~$\sigma$ rms noise in the final
  1.3~mm continuum map for a temperature of 20~K. Data for a high-mass
  CygnusX-N53 \citep{bontemps2010:cygX} and low
  mass protocluster  Oph-B2 \citep{andre2007} is shown for
  comparison. The masses are derived from dust continuum for all the
  cores. \textit{Panel (B)}:
  virial mass (scaled by mean virial parameter 0.3) derived from \DAMM\ data versus separation. Data from
  two low mass clumps, Oph-B2 \citep{andre2007}, and IRAS 19175-4/5
  \citep{beuther09_lowmass} is also shown for
  comparison. The masses correspond to virial masses derived from  $\rm N_2H^+$  1--0 observations. 
\textit{Panel (C)}: Same as panel B with the
  corresponding thermal Jeans mass and length for the respective core
  density, and temperatures. \textit{Panel (D)}: Same as panel B with
  the corresponding turbulent Jeans mass and length for the respective
  core density, and line width.}
\label{fig:fragmn}
\end{figure}

This data on cloud fragmentation can be directly compared to
predictions from Jeans-type cloud fragmentation
(\citet{kippenhahn1990:stel_str}; their Sect.\ 26.3). In this model, an
initially homogeneous state of given characteristic $\rm H_2$ particle
density and velocity dispersion, $n_{\rm char}$ and
$\sigma_{\rm char}$, will fragment into features of mass
\begin{equation}
M_{\rm J} = \varrho \, \ell_{\rm J}^3 =
1.578 \, M_{\sun} \,
\left( \frac{\sigma_{\rm char}}{0.188 ~ \rm km \, s^{-1}} \right)^3 \, 
\left( \frac{n_{\rm char}}{10^5 ~ \rm cm^{-3}} \right)^{-1/2} \, ,
\label{eq:jeans-mass}
\end{equation}
which are separated by a distance
\begin{equation}
\ell_{\rm J} =
\left( \frac{\pi}{G \varrho} \right)^{1/2} \, \sigma_{\rm char} =
0.06 ~ {\rm pc} \,
\left( \frac{\sigma_{\rm char}}{0.188 ~ \rm km \, s^{-1}} \right) \, 
\left( \frac{n_{\rm char}}{10^5 ~ \rm cm^{-3}} \right)^{-1/2}
\label{eq:jeans-length}
\end{equation}
($\varrho$ is the mass density implied by $n_{\rm char}$).  By
comparing the observed masses of cores, and separations between them,
to the predicted Jeans-values, we can thus assess whether
Jeans-type fragmentation governs the structure formation in our target
regions. This is shown in Fig.\ \ref{fig:fragmn}.

The densities and velocity dispersions substituted in Eqs.\
(\ref{eq:jeans-mass}, \ref{eq:jeans-length}) have to be chosen
carefully. For a given core, the density will lie between
the one of the actual dense core, and the one of the SCUBA-detected
envelope. To bracket the
reasonable parameter range, we adopt either
$n_{\rm char} \to n_{\rm core}$ or $n_{\rm char} \to n_{\rm env}$ in
the calculations below. We derive the envelope and core densities from SCUBA 850\,$\mu$m and PdBI data respectively as, $n_{\rm char}= \frac{\rm Mass}{\frac{4}{3} \,\pi \, {\rm radius}^3}$. Similarly, we use either the thermal velocity
dispersion of the mean free particle, 
$\sigma_{\rm char} \to \sigma_{\rm th}^{\rm mean}$ (using a mean
molecular weight of 2.33; \citealt{kauffmann2008}), or we also include
the impact of non-thermal motions,
$\sigma_{\rm char} \to
[(\sigma_{\rm th}^{\rm mean})^2 + (\sigma_{\rm nt})^2]^{1/2}$. Depending on
whether the core or envelope density is used, we adjust the
non-thermal velocity dispersion, $\sigma_{\rm nt}$, to the value for
the core or its envelope.

\subsubsection{Analysis}
Fragmentation is best studied at the highest possible
resolution. Here, we rely on our 1.3~mm PdBI maps. Ideally, we
would base our analysis on dust emission data, which provides a fairly
reliable probe of mass reservoirs. This is shown in panel A of Fig.\
\ref{fig:fragmn}. However, only few cores are detected in dust
emission at 1.3~mm wavelength. Therefore, we use the separations and
virial masses of \DAMM{} cores in the following. Here, core separation
  is defined as the distance to the nearest neighbor as seen
  in projection on the sky. Since we find
  that virial
  masses are always much smaller than the dust masses, we scaled
  the virial mass by the mean virial parameter (0.3) to reflect the actual
  core masses.  They are presented in
Fig.\ \ref{fig:fragmn}B. To guide the reader through the following discussion,
  the mean observed core size, mass, and density is given by R$_{\rm
    eff}=0.14(0.04)$~parsecs, $n_{\rm core}=1.3(8.3)\times10^5$~\percc\  and
  M$_{\rm vir}=110(33)$~\Msol\ for G29.96e(G35.20w).

First consider the case of thermal fragmentation at envelope density
($n_{\rm char} \to n_{\rm env}$,
$\sigma_{\rm char} \to \sigma_{\rm th}^{\rm mean}$), indicated by
pentagons in panel C of Fig.\ \ref{fig:fragmn}. In G29.96e, the predicted
mass for this case falls short of observed values by more than a factor
10. In separation, the discrepancy is up to a factor 10. Similar
maximum discrepancies are seen in G35.20w.  Thus, the structure of G29.96e and G35.20w is not
well described by thermal fragmentation of the envelope. A very similar discrepancy prevails even if we assume
core densities ($n_{\rm char} \to n_{\rm core}$,
$\sigma_{\rm char} \to \sigma_{\rm th}^{\rm mean}$; triangles in Fig.\
\ref{fig:fragmn}C).

Turbulent Jeans fragmentation at envelope density
($n_{\rm char} \to n_{\rm env}$, $\sigma_{\rm char} \to [(\sigma_{\rm
th}^{\rm mean})^2 + (\sigma_{\rm nt}^{\rm env})^2]^{1/2}$, pentagons
in Fig.\ \ref{fig:fragmn}D) predicts masses and separations that is
inconsistent with observations as seen for thermal
fragmentation. The observed values are always lower
than the model values. Finally, turbulent Jeans fragmentation at core
density ($n_{\rm char} \to n_{\rm core}$,
$\sigma_{\rm char} \to [(\sigma_{\rm th}^{\rm mean})^2 +
(\sigma_{\rm nt}^{\rm env})^2]^{1/2}$; triangles in Fig.\
\ref{fig:fragmn}D) yields masses and separation predictions that are
on average within a factor 3 of the observations. Among the models explored
here, this is the best fit to the data. 

Thus, the observed masses and separations of the cores are best
understood if they result from turbulent Jeans fragmentation. An observable predicted by
turbulent fragmentation models is that clump separation scales should scale with mass \citep{padoan2002}.
We also note that gravitational fragmentation models also claim to make a similar prediction (see Sect.5.2 of \citealt{bonnell2007} for a discussion).
Interestingly, our observations lie closer to those predicted for core densities. Thus, it seems plausible that
fragmentation occurred closer to core than envelope densities and velocity dispersions.

\subsubsection{Previous results}
Numerous papers have addressed cloud fragmentation over the last
decade. Figures \ref{fig:fragmn}A and \ref{fig:fragmn}B present, for
example, data that are representative of low to intermediate to higher
mass protoclusters. Oph-B2 is a well studied low mass protocluster region
in the Ophiuchus molecular cloud complex with a total mass of
$42 \, M_{\sun}$ \citep{motte1998:ophiuchus}. $\rm N_2H^+$ observation
imply subsonic to transonic turbulence \citep{andre2007}.
\citet{beuther09_lowmass} study the fragmentation of two clumps of low
to intermediate mass, IRAS 19175-4 and 5 (combined single-dish mass of
$87 \, M_{\sun}$). The velocity dispersions are transonic to
supersonic in these sources, just as seen in our targets (Sect.\ref{subsec:linewidth}). Also the massive clumps mapped by \citet{zhang09_g28} belong to
this latter velocity domain. Ideally, one would want to perform a fragmentation study such as ours on entire massive star-forming complexes like Cygnus-X \citep{motte2007}. Recently, \citet{bontemps2010:cygX} have studied six individual massive dense cores within this complex in dust continuum at high angular resolution. Bontemps et al.\  note that the core size, and separation
suggest densities much higher than in a turbulence regulated star
formation scenario proposed by \citet{mckee02:100000yrs}. It would be
interesting to combine the continuum data with high angular resolution
spectral line data and compare the results with our study. Since this
information is not yet available, we derived the minimum separation
for one of their massive dense cores CygnusX-53. CygnusX-53 was chosen
because their 1.3~mm PdBI continuum observations shows the highest fragmentation toward this core. Comparison with our results
is shown in Figure \ref{fig:fragmn}A. Clearly, for the same spatial
scales, a few cores in CygnusX-N53 have an order of magnitude higher mass than in Oph-B2, the low mass protocluster, in agreement with our naive
expectation. Whether fragmentation is turbulence regulated is yet to
be answered.

\citet{motte1998:ophiuchus} and \citet{beuther09_lowmass} both
conclude that their separation data are consistent with thermal core
fragmentation ($n_{\rm char} \to n_{\rm core}$,
$\sigma_{\rm char} \to \sigma_{\rm th}^{\rm mean}$). In contrast,
\citet{zhang09_g28} find that thermal envelope fragmentation
($n_{\rm char} \to n_{\rm env}$,
$\sigma_{\rm char} \to \sigma_{\rm th}^{\rm mean}$) fails to
explain their mass and separation observations. They suggest that
turbulence, magnetic fields, or both, may influence fragmentation. 
Thus, our conclusions are more comparable with the observations by
\citet{zhang09_g28}. This is not necessarily a surprise, given that our study and 
\citet{zhang09_g28}  explore similar regions, while
\citet{motte1998:ophiuchus} and \citet{beuther09_lowmass} study clumps
of much lower mass and turbulence.

\subsubsection{Fragmentation sequences \& cascades}
Intriguingly, all of the fragmentation data plotted in Fig.\
\ref{fig:fragmn} lie along a common sequence, independent of the
nature of the region studied. Mass-separation data for clouds forming
stars of lower mass fall at the end with small masses and 
separations, while observations for turbulent regions forming higher
mass stars constitute the opposite end of the sequence. Here we lack
the space to discuss the nature of this sequence comprehensively;
resolution and intensity detection thresholds may well conspire to
yield such an apparent correlation. It may be instructive to explore
this finding in future studies.\medskip

\noindent Another result of the fragmentation analysis is that all studies are
consistent with fragmentation occurring at densities and velocity
dispersion similar to the present-day values for the cores. Most
likely, $n_{\rm char} \to n_{\rm core}$ yields good matches because
the telescope resolution highlights a particular spatial scale. At
that scale, $n_{\rm core}$ is a characteristic density, simply because
of the way cores are extracted. Fragmentation probably occurred at a
slightly larger spatial scale of slightly lower density. If true,
$n_{\rm char} \to n_{\rm core}$ in fragmentation calculations will
thus naturally yield a good description for the emission that is just
resolved.

This argument only works if fragmentation  continuously occurs at
any given spatial scale and density (perhaps with a limit at small
scales and high densities). For any given resolution, apparently
monolithic objects would break up into smaller fragments, if studied
with sufficient resolution. We can detect fragmentation down to $\sim$ twice the beam size (0.02pc for G35.20w, or 0.07pc for G29.96e) 
but any fragmentation on scales smaller than this would go undetected. In this situation, the fragment masses
derived here (or in any other study) only presents a snapshot of the
fragmentation cascade taken at a particular scale. Then, they bear no
obvious relation to the mass of the star eventually forming from the
cloud. This only breaks down at the bottom of the cascade, once
fragments with a mass $< M_{\rm J}$ are resolved. 

%TABLE 11.CORE MASSES FROM DUST CONTINUUM
\begin{table*}
\begin{minipage}{\textwidth}
\begin{center}
  \caption{Cores identified from the 1.3~mm and 3.5~mm continuum
    emission. Columns are the clump name (for 1.3mm cores only),
    offset position relative to the map center, peak flux, total flux,
    the effective radius after deconvolution with the beam of the mm
    core, total gas mass for a temperature T$_{\rm kin}$ estimated
    from \AMM\ observations, peak ${\rm H_2}$ column density, average
    gas density over R$_{\rm eff}$, total mass, virial mass from
    \DAMM\ observations, the alpha parameter as defined in
    Eq.\ref{eq:alpha}, and an aperture mass for an aperture size of
    37000 AU. 37000~AU corresponds to the beam of the more distant
    source G29.96e. The aperture mass is underestimated up to a factor
    2 when aperture size is very similar to the
    beam. \label{tab:pdbi_cont_mass} }
\vspace*{2mm}

\begin{tabular}{ccrrrccccccc}
\hline
Core & core offsets & S$_{\rm peak}$ & S$_{\rm total}$ & R$_{\rm eff}$ & T$_{\rm kin}$ & N$_{\rm H_2}$    &   $\rho$ & M$_{\rm total}$ &  M$_{\rm vir}$ & $\alpha$ & M$_{\rm aperture}$  \\
& arcsec     & Jy/beam          &   Jy            & parsecs       & (K)           & (10$^{23}$ \cmsq) &  (10$^{5}$ \percc)  & \Msol          &    \Msol       &             & \Msol         \\

\hline
 &           &         &          &        &   & {\bf G29.96e-1.3mm}         &         &          &          &           &          \\       
mm1 & ( 2,  10)    & 0.052   & 0.293    &  0.16     &   23.        &    4.   &     2.   &   291.   &   90.   &    0.3   &   135. \\            
mm2 & (-2,   0)    & 0.026   & 0.200    &  0.17     &   16.$^a$    &    4.   &     2.   &   320.   &   55.   &    0.2   &   107. \\            
mm3 & ( 4,  16)    & 0.024   & 0.055    &  0.07     &   16.$^a$    &    4.   &     9.   &    88.   &   19.   &    0.2   &    90. \\            
mm4 & ( 4,  20)    & 0.020   & 0.051    &  0.08     &   16.$^a$    &    3.   &     6.   &    82.   &   13.   &    0.2   &    83. \\            
mm5 & ( 2, -14)$^b$& 0.018   & 0.054    &  0.09     &   18.        &    2.   &     4.   &    74.   &   --    &    --    &    57. \\            
mm6 & (-1,  18)$^c$& 0.014   & 0.036    &  0.07     &   16.$^a$    &    2.   &     6.   &    58.   &   --    &    --   &    45. \\            
mm7 & (-1, -24)    & 0.012   & 0.028    &  0.07     &   13.        &    2.   &     6.   &    60.   &   51.   &    0.8   &    32. \\            
&             &         &          &        &   & {\bf G29.96e-3.5mm}        &          &          &         &          &          \\  
    & ( 2,  10)    & 0.010   & 0.026    &  0.24     &   20.        &    5.   &     2.   &   836.   &  110.   &    0.1   &   225. \\           
    & (-1,   2)    & 0.007   & 0.017    &  0.21     &   20.        &    3.   &     2.   &   547.   &   64.   &    0.1   &   161. \\           
    & ( 2, -15)    & 0.003   & 0.005    &  0.14     &   17.        &    2.   &     2.   &   193.   &   42.   &    0.2   &    77. \\           
&             &         &          &         &  & {\bf G35.20w-1.3mm}        &          &          &         &          &          \\     
mm1 & ( 4,  -1)    & 0.044   & 0.090    &  0.03     & 	 22.        &    5.   &    24.   &    18.   &    7.   &    0.4   &    17. \\  
mm2 & (16,   3)    & 0.030   & 0.059    &  0.03     & 	 21.        &    3.   &    17.   &    13.   &    9.   &    0.7   &    22. \\  
mm3 & (13,   2)$^b$& 0.028   & 0.051    &  0.03     & 	 21.        &    3.   &    14.   &    11.   &   --    &    --   &    18. \\  
mm4 & (6,  12))$^{c,d}$& 0.013   & 0.012    &  0.01     & 	 10.        &    5.   &   265.   &     8.   &   --    &    --   &     6. \\  
 &            &         &          &         &  & {\bf G35.20w-3.5mm}        &          &          &         &          &          \\     
    & ( 4,  -1)    & 0.017   & 0.078    &  0.15     &   20.        &    9.   &     5.   &   490.   &  107.   &    0.2   &   264. \\                  
    & (15,   2)    & 0.011   & 0.046    &  0.13     &   19.        &    6.   &     5.   &   306.   &  145.   &    0.5   &   206. \\                  
    & (-13, -11)   & 0.003   & 0.009    &  0.08     &   16.$^a$    &    2.   &     5.   &    78.   &    8.   &    0.1   &    48. \\                  
    & (-15,  12)   & 0.003   & 0.005    &  0.06     &   16. $^a$   &    2.   &     6.   &    40.   &   13.   &    0.3   &    32. \\                  
\hline
\end{tabular}
\end{center}

\medskip
Comments: $^a$ Mass estimates based on an assumed temperature of ~16
K, since measured value has a very high error  ($>30\%$).\\
 $^b$ \DAMM\ spectra at these offsets show clearly more than one velocity component. We did not consider these positions because of the following, (i) poor two-component HFS fit in CLASS in one case, (ii) difficulty of associating a particular component to the dust.\\
 $^c$ No \DAMM\ detection.
 $^d$ mm4 is barely resolved with peak emission just above
   5$\sigma$. Given the biases in a typical interferometer image, this
 might be spurious.
\end{minipage}
\end{table*}

\section{Models of  star formation: stability \&
  accretion\label{sec:theory}}
In recent years, two models for the formation of massive stars have
proven particularly influential: (\textit{i}) the monolithic collapse
of massive cloud fragments supported by supersonic turbulent pressure
(\citealt{mckee02:100000yrs}; essentially like
\citealt{shu1977:self-sim_collapse}, with scaled-up velocity
dispersions); and (\textit{ii}) rapid growth of many cores of
initially very low mass via competitive accretion from a common
massive mass reservoir (\citealt{zinnecker82:orion}; Prediction of the
protostellar mass spectrum in the Orion near-infrared cluster;
\citealt{bonnell2001a, bonnell2004, bonnell2007}). Some of our data
are suited to test the one or the other of these two scenarios.

Beyond this, we derive estimates of various key properties of out
target clouds. This includes their stability against collapse,
magnetic fields, and rates for mass accretion \emph{onto} the cores.

\subsection{Turbulent and magnetic cloud support\label{subsec:vir}}
Random motions in excess of the thermal ones can contribute to the
support of an object against self-gravity
\citep{bertoldi1992:pr_conf_cores}. The impact of such turbulent
motions is captured by the virial mass and ratio (Eqs.\
\ref{eq:viriall-mass}, \ref{eq:alpha}). In these, the adopted velocity
dispersions reflect thermal as well as further random motions. A cloud
fulfilling the stability criteria of \citet{ebert1955:be-spheres} and
\citet{bonnor1956:be-spheres}, for example, has $\alpha > 2.06$
\citep{bertoldi1992:pr_conf_cores}. The exact stability threshold with
respect to collapse depends on the nature of the adopted equation of
state. For reasonable assumptions, $\alpha \ll 1$ implies a lack of
pressure support against collapse, hence subvirial.

As shown in tables \ref{tab:nh2d_line_parameter} and
\ref{tab:pdbi_cont_mass}, $\alpha \ll 1$ holds for all cores and
clumps identified in our target
regions. Table~\ref{tab:pdbi_cont_mass} gives the $\alpha$ values for
dust and virial mass computed from the 3.5~mm PdBI continuum and line
(\DAMM) data corresponding to the dust cores. The mean (median)
$\alpha$ for the cores is 0.3 (0.2).  Similarly, the \DAMM\ cores
identified with CLUMPFIND and listed in
Table~\ref{tab:nh2d_line_parameter} have also a mean alpha of
$0.3$. We do the same estimate for the clump scale for G29.96e and
G35.20w since we have the SCUBA data.  Comparing the virial mass
derived based on average line width from \CO\ 2--1 IRAM 30m
observations to that of the clump mass from SCUBA data for a
$\sim$~38\arcsec\ aperture around the brightest peak in G29.96e and
G35.20w, we estimate alpha to be 0.3 and 0.2 respectively. Here we have used the relatively low density gas \CO\ line widths as
opposed to \DAMM\ as \CO\ is more representative of envelope
properties.  We thus conclude that the observed supersonic turbulent
motions, in combination with the (negligible) thermal support, are not
sufficient to prevent the cores from collapsing.

Our assessment that $\alpha{}\ll{}1$ is robust with respect
to observational uncertainties. Equation (\ref{eq:alpha}) shows that
$\alpha{}\propto{}\sigma^2{}R/M$. The uncertainties on the velocity
dispersion, $\sigma$, are negligible. The radius, $R$, has no
uncertainty, since it is a freely chosen parameter (as long as
$\sigma$ and $M$ are measured for the chosen $R$). The mass is ---
because of our limited knowledge of dust opacities and temperatures
--- probably uncertain by a factor $\sim{}2$. This implies a similar
uncertainty for $\alpha$. Since our typical observation is
$\alpha{}\lesssim{}0.3$, this uncertainty still robustly implies
$\alpha{}\ll{}1$. Furthermore, the spatial filtering implies actual masses
exceeding the observed ones, and thus virial parameters lower than
the derived ones. Large distance errors, $\delta{}d$, are needed to
yield large errors in $\alpha$:
$(\alpha+\delta{}\alpha{})/\alpha=[(d+\delta{}d)/d]^{-1}$. Distance
errors larger than a factor 2 are thus needed to significantly affect
$\alpha$. This seems unlikely, in particular given the parallax-based
result for G35w.

The observation $\alpha \ll 1$ does not imply that the cloud fragments
are collapsing. In particular, there might be additional support from
magnetic fields, for which we did not account in the above calculation.
Following the virial energy argument used in the initial definition of
$M_{\rm vir}$ (see \citealt{bertoldi1992:pr_conf_cores} for details),
we can define a magnetic virial mass,
\begin{equation}
M_{B, \rm vir} = \frac{5 R}{G} \,
\left (\sigma^2 + \frac{1}{6} \sigma_{\rm A}^2 \right) \, ,
\end{equation}
and a magnetic virial ratio,
\begin{equation}
\alpha_B = \frac{M_{B, \rm vir}}{M} \, ,
\label{eq:alpha_m}
\end{equation}
where
\begin{equation}
\sigma_{\rm A} = \frac{B}{(\mu_0 \varrho)^{1/2}}
\label{eq:bfield}
\end{equation}
is the Alfv\'en velocity for given magnetic field, $B$ (in SI;
$\mu_0$ is the permeability of free space). This approach captures the
basic aspects of magnetic cloud support, and is well-suited to
order-of-magnitude estimates. Researchers requiring a higher degree of
accuracy, though, may wish to employ more accurate schemes (see
Sect.\ 2.3 of \citealt{bertoldi1992:pr_conf_cores}).

The magnetic field necessary to support the fragments can be estimated from
Eqs.\ (\ref{eq:alpha_m}, \ref{eq:bfield}) by requiring $\alpha_B = 1$.
For the entire clump, we get $670 ~ \rm \mu G$ (G29.96e), and
$1650 ~ \rm \mu G$ (G35.20w). For the cores in both filaments, we find
that on average $\sim 1 ~ \rm mG$ field is needed to fulfill
$\alpha_B = 1$. Such field strengths may seem high and unrealistic,
in light of recent measurements for low mass dense cores. However,
observations spanning a large range in densities have revealed a
field-density relation, such that $B \propto n^{1/2}$
\citep{troland86, padoan1999:super_alfvenic_model, vlemmings08_aspc}.
With a lot of upper limits, particularly at the cold core densities,
this relation has a significant amount of scatter. Even so, it is
interesting to note that the limits on magnetic fields derived by us
are not grossly inconsistent with the proposed relation. Specifically,
our results for field strengths are not inconsistent with the
compilation by \citet{padoan1999:super_alfvenic_model}, examined at
densities $\sim 10^5 ~ \rm cm^{-3}$. However, a preliminary
analysis of our own Zeeman and polarization observations towards high-mass cores are
inconsistent with a high magnetic field, which would be relatively
easy to detect (Pillai et al., work in progress).\medskip

\noindent Our observations of $\alpha \ll 1$ render the
\citet{mckee02:100000yrs} model quantitatively not applicable to our
target clouds. To see this, consider the basic assumptions of their
model: cloud fragments are modeled as equilibria supported by
turbulent random motions. Our targets do not bear any similarity with
such equilibria, though: these would have $\alpha \gtrsim 1$ (e.g.,
\citealt{mckee2003:turbulence}), while our observations show
$\alpha \ll 1$. In detail, our observations do thus not rule out
monolithic collapse. However, the quantitative description by
\citet{mckee02:100000yrs} cannot apply here.

Global infall motions, as suggested by our $\alpha \ll 1$
observations, are a common feature in many models of competitive
accretion. In these models, the whole clump undergoes global collapse,
and the gas accretion and the protocluster evolution occurs on the
global dynamical timescale. Recently, \citet{krumholz05:comp_accr} did
actually use $\alpha$ to rule out models with competitive accretion.
They assert an apparent lack of observational evidence for
$\alpha \ll 1$---a gap now filled by our data. In fact, our
observations are not the first to suggest a possible global
infall. Two recent studies towards low and intermediate mass star
forming regions find evidence of such large-scale collapse
(\citealt{andre2007}, \citealt{peretto2007}). Also, note that a
preliminary numerical model by \citet{vazquez-semadeni2008} indicates
that physical properties in high-mass star-forming regions can be
reproduced by starting with such a global infall.

\subsection{Accretion rates onto cores and stars}
The formation of massive stars inevitably requires high accretion
rates, $\dot{M}$. A fundamental constraint is that a star of
mass $M_{\star}$ must accrete all of its mass in one accretion time
scale, $\tau_{\rm accr}$. Then,
$\dot{M}_{\star}^{\rm time} = M_{\star} / \tau_{\rm accr}$, respectively
\begin{equation}
\dot{M}_{\star}^{\rm time} = 10^{-4} {M_{\sun} ~ {\rm yr^{-1}}}
\left( \frac{M_{\star}}{10 ~M_{\sun}} \right) \,
\left( \frac{\tau_{\rm accr}}{10^5~\rm{yr}} \right)^{-1} \, .
\label{eq:acc}
\end{equation}
A normalization of $\tau_{\rm accr}$ to $\lesssim 10^5 ~ \rm yr$ is
implied by several lines of argument. Recent Spitzer surveys of solar
neighborhood clouds yield accretion timescales
$\sim 5 \cdot 10^5 ~ \rm yr$ for low-mass stars
\citep{evans09_c2d}. Given the shorter free-fall times for the more
massive cores considered here, one may thus expect shorter timescales
for massive stars. Furthermore, theoretical models of the collapse of
massive dense cores imply durations $\sim 10^5 ~ \rm yr$ \citep{mckee02:100000yrs}. 
Observational constraints from source
counts for different evolutionary stages actually imply
$3 \cdot 10^4 ~ \rm yr$ (\citealt{motte2007}, \citealt{parsons09}). The
duration of the accretion phase thus robustly implies
$\dot{M}_{\star}^{\rm time} \gtrsim 10^{-4} {M_{\sun} ~ {\rm yr^{-1}}}$ for
stars more massive than $10 \, M_{\odot}$.

A particular constraint for spherical accretion on massive stars
results from the ``accretion luminosity problem'': the momentum of the
accreting matter has to overcome the star's radiation pressure
\citep{wolfire1987}. Following \citet{jijina1996}, we estimate that a
rate exceeding
\begin{equation}
\dot{M}_{\star}^{\rm lumi} = 1.2 \times 10^{-6} {M_\odot ~ {\rm yr^{-1}}}
\left( \frac{M_\star}{10 ~ M_\odot} \right)^{-1/2} \,
\left( \frac{L_\star + L_{\rm accr}}{5 \cdot 10^3 ~ L_\odot} \right)^{5/4}
\label{eq:acc-lumi}
\end{equation}
is needed to overcome the radiation pressure. In this, the combined
luminosity of star and accretion, $L_{\star} + L_{\rm accr}$, is
normalized to regular observed luminosities. Note that
$\dot{M}_{\star}^{\rm lumi} < \dot{M}_{\star}^{\rm time}$ for the case considered
here. Thus, the accretion luminosity problem does not imply higher
rates than required by basic lifetime arguments. Thus,
$\dot{M}_{\star} \approx \dot{M}_{\star}^{\rm time}$.

Direct observations of accretion flows in \HII\ regions actually imply
even higher accretion rates $\sim 10^{-3} {M_\odot ~ {\rm yr^{-1}}}$
\citep{keto2006}. This is in line with observations of nearby young
high-mass stars (e.g., \citealt{cesaroni05}). It is not clear, though,
how representative these values are.\medskip

\noindent Equations (\ref{eq:acc}, \ref{eq:acc-lumi}) imply high
\emph{stellar} accretion rates (i.e., onto the stellar surface). These
rates have to be supplied by the stellar environment, such as the
cores identified here. Thus, it is interesting to also consider
\emph{core} accretion rates, i.e., the rates by which cores (or any
other environment of a star) gains mass from its surroundings. Both
these rates are considered below.

\subsubsection{Accretion of mass onto stars}
First, let us consider the rate at which mass accretes onto the
central star. A very basic accretion model is the monolithic collapse
of the entire core of mass $M$ during the free--fall timescale,
$\tau_{\rm{}ff}=(3\pi{}/[32G\langle{}\varrho{}\rangle])^{1/2}$ ($G$ is
the constant of gravity, and $\langle{}\varrho{}\rangle$ the mean
density), so that $\dot{M}_{\star}^{\rm{}ff}=M/\tau_{\rm{}ff}$. In
practice,
$\tau_{\rm{}ff}=9.8\times{}10^4~{\rm{}yr}\,(\langle{}n\rangle{}/10^5~{\rm{}cm^{-3}})^{-1/2}$. We
find mean free--fall accretion rates of
$0.6\times{}10^{-3}\,M_{\sun}\,\rm{}yr^{-1}$ and 
$3\times{}10^{-3}\,M_{\sun}\,\rm{}yr^{-1}$ for G35.20w and G29.96e,
respectively. This is well within the range required for massive star
formation.\medskip

\noindent \citet{mckee02:100000yrs, mckee2003:turbulence} present a
more elaborate model of such a monolithic collapse. As an initial
condition, the model chooses an equilibrium core supported by
non-thermal pressure described by a velocity dispersion, $\sigma$, and
a polytropic exponent $\gamma_P$. Then, the core undergoes the
classical inside-out collapse, resulting into accretion onto a star
forming in the core's center. Re-evaluation of the
\citeauthor{mckee02:100000yrs} calculations shows that the maximum
accretion rate from a configuration that is initially in equilibrium
is found for $\gamma_P = 1.06$,
\begin{equation}
\dot{M}_{\star}^{\rm eq} \le 2.3 \times 10^{-4} \, M_{\sun} \,
\left( \frac{\sigma}{\rm km \, s^{-1}} \right)^3 \, .
\label{eq:accretion-coll_equil}
\end{equation}
Unfortunately, it is not clear what values are to be substituted for
$\sigma$. If we use the greatest velocity dispersions,
$(8\ln[2])^{-1/2}\,{}\Delta{}v$, resulting from the observations
reported in Table \ref{tab:nh2d_line_parameter}, then we derive
implied accretion rates of up to
$0.6\times{}10^{-4}\,M_{\sun}\,\rm{}yr^{-1}$ and
$1.1\times{}10^{-4}\,M_{\sun}\,\rm{}yr^{-1}$ for G29.96e and G35.20w,
respectively. This is about consistent with the rate of
$10^{-4}\,M_{\sun}\,\rm{}yr^{-1}$ implied by Eq.\
(\ref{eq:acc}). Also, we find these high rates primarily towards the
dust emission peaks where massive stars are observed to
form. \emph{But this also suggests that the velocity dispersions are
  increased by stellar feedback and are not representative of the
  initial conditions to be substituted in Eq.\
  (\ref{eq:accretion-coll_equil}).}

On average, the observations in Table \ref{tab:nh2d_line_parameter}
imply
$\langle{}\dot{M}_{\star}\rangle{}\le{}3\times{}10^{-5}\,M_{\sun}\,\rm{}yr^{-1}$.
If representative of the initial state, this is too low for massive
star formation. Also, recall that $\alpha{}\ll{}1$, which 
renders the \citet{mckee02:100000yrs, mckee2003:turbulence} model not
applicable.  This observation does, however, not challenge general
models of monolithic collapse, which may well obey other accretion
laws than explored here.

\subsubsection{Accretion of mass onto cores}
We may also consider the mass evolution of the reservoir (the
``core'') from which a star forms. In our study, this reservoir does
probably about correspond to the cores extracted from the
interferometer maps. It depletes in mass via accretion onto the
star. But the reservoir may also gain mass by accretion onto
itself. Here, we derive a new idealized model for this process.\medskip

\noindent A very basic model of accretion onto cores can be
established when idealizing a core as a sphere with radius $R$ that
coasts at speed $v$ through a medium of density $\varrho$ and
  velocity dispersion $\sigma_{\rm{}env}$. Provided collisions with
this sphere are inelastic, the sphere-environment collisions due to
the sphere's motion, as well as the drift of gas particles
  towards the sphere due to random motion, will collect mass at a
rate
\begin{equation}
\dot{M}_{\rm{}core} = \pi R^2 \varrho
  \left[
    \sigma + \frac{\pi}{2} v +
    \int_0^{1/2} \left| \sigma - v \, \cos(\vartheta) \right| \,
                \sin(\vartheta) \, {\rm{}d} \vartheta
  \right] \, .
\label{eq:accretion-general}
\end{equation}
This equation can be derived by separately considering the gas
  particles facing the ``front side'' and the ``back side'' of the
  moving sphere. On the front side, for the first half of particles
  with random motion towards the core, the mean core--gas relative
  velocity is $v_{\rm{}rel}=v\,{}\cos(\vartheta)+\sigma$. For the
  other half, $v_{\rm{}rel}=v\,{}\cos(\vartheta)-\sigma$, where
  $\vartheta$ is the position angle of a particle with respect to the
  core's direction of motion. Similar relations hold for the gas on
  the back side. At every position on the sphere, mass contained in
  one of the four populations of gas particles (i.e., two gas motion
  directions per hemisphere) hits the surface at a rate
  ${\rm{}d}\dot{M}_{\rm{}core}=(\varrho/2)\,v_{\rm{}rel}\,{\rm{}d}A$,
  as long as $v_{\rm{}rel}>0$ for the gas population considered. In
  this, ${\rm{}d}A=2\pi{}R^2\,\sin(\vartheta)\,{\rm{}d}\vartheta$ is
  the surface element. Summation over the populations, and integration
  over all angles where $v_{\rm{}rel}>0$, then yields Eq.\
  (\ref{eq:accretion-general}).

It is convenient to examine two extreme cases, in which Eq.\
  (\ref{eq:accretion-general}) assumes are more simple form. First,
  let us consider the case of cores moving through their environment
  at a very high speed, $v\gg{}\sigma$. Then, the rate at which the
  core accretes mass that ``flies by'' becomes $\varrho{}Av$, or
\begin{eqnarray}
\dot{M}_{\rm core}^{\rm{}fly-by} & = &
2.2 \times 10^{-5} \, M_{\sun} \, {\rm yr}^{-1} \, \nonumber \\
& & \quad \quad \left( \frac{n}{10^4 ~ \rm cm^{-3}} \right) \,
\left( \frac{v}{\rm km \, s^{-1}} \right) \,
\left( \frac{R}{0.1 ~ \rm pc} \right)^2 \, ,
\end{eqnarray}
where $A$ is the cross section. To estimate the relative
core-envelope motion, we may assume that the ensemble of cores rests
with respect to the environment. Then, the dispersion of the core-core
relative motions, $\sigma_{\rm core-core}$, gives the mean
core-environment relative motion along the line of sight. In 3-dimension, the relative motion is larger by a factor $3^{1/2}$,
i.e. $v=3^{1/2} \times \sigma_{\rm core-core}$. Table \ref{tab:acc}
presents some rate estimates calculated for these assumptions. The
environmental density is estimated based on SCUBA data (estimate from
\citealt{curran2004} for G35.20w). The G29.96e is broken up into
several parts, to reflect the variety of cloud conditions in this
region. We find fly-by accretion rates in the range $0.4 ~ {\rm to} ~
5 \times 10^{-4} \, M_{\sun} ~ \rm yr^{-1}$.

Second, let us explore the situation of vanishing
  core-environment motion, $v\ll{}\sigma$. This ``turbulent''
  accretion is only due to random motion of particles towards the
  sphere. The rate becomes $\varrho{}A\sigma_{\rm{}env}/2$, in which
  $A$ is the core's surface, or
\begin{eqnarray}
\dot{M}_{\rm core}^{\rm{}turb} & = &
4.4 \times 10^{-5} \, M_{\sun} \, {\rm yr}^{-1} \, \nonumber \\
& & \quad \quad \left( \frac{n}{10^4 ~ \rm cm^{-3}} \right) \,
\left( \frac{\sigma_{\rm{}env}}{\rm km \, s^{-1}} \right) \,
\left( \frac{R}{0.1 ~ \rm pc} \right)^2 \, .
\end{eqnarray}
We gauge $\sigma_{\rm{}env}$ by setting it equal to the velocity
dispersion in the $\rm{}C^{18}O$ 2--1 transition, as observed with the
IRAM 30m--telescope towards the phase center of the PdBI map. Values
are given in Table \ref{tab:acc}. Rates $\ge{}3\times{}10^{-4}$~\Msol{}
are found.\medskip

% TABLE XX. Accretion rate 
\begin{table*}
\begin{center}
  \caption{Core accretion rates. The columns are, (i) the region of the
    filament, (ii) number of cores for that region, (iii) mean radius of the
    cores, (iv) mean velocity dispersion from PdBI \DAMM\, (v) mean
    $\rm{}C^{18}O$ envelope velocity dispersion, (vi) mean
    protocluster density (SCUBA 850$\mu$m), (vii) mean fly--by accretion
    rate, and (viii) mean turbulent accretion rate. \label{tab:acc}}
\vspace{1em}
\begin{tabular}{lcccccccc}
\hline
Source  & $N_{\rm cores}$ & $R$ & $\sigma_{\rm core-core}$ &
$\sigma_{\rm{}env}$ & $n$ &
 $\langle \dot{M}_{\rm core}^{\rm{}fly-by} \rangle$ &
 $\langle \dot{M}_{\rm core}^{\rm{}turb} \rangle$\\
 &  & pc & \kms & \kms & $10^4$~\percc &  $10^{-4}$~\Msol$\rm yr^{-1}$ &
$10^{-4}$~\Msol$\rm yr^{-1}$  \\ \hline
{\bfseries \itshape G29.96e} \\
North filament & 4  & 0.22 & 0.5 & 1.4 & 5   & 4.6 & 14.5  \\
South filament & 3  & 0.15 & 0.2 & 1.4 & 2.5 & 0.4 &  3.4 \\
{\bfseries \itshape G35.20w} \\
Whole filament & 13 & 0.06 & 0.5 & 1.1 & 30 & 2.3 & 5.2 \\
\hline
\end{tabular}
\end{center}
\end{table*}

\noindent This model of core accretion is very simplistic;
for example, if collisions were truly inelastic, the inhibition of
rebound after particle collisions would imply the strict absence of
kinetic pressure. Still, our toy model shows that a dense
core is in steady contact with a large mass reservoir, from which it
could at least in principle collect mass at a significant rate.

Thus, the accretion rate onto the core might be similar to the
accretion rate onto the massive stars, i.e.\ $\dot{M}_{\rm core} \sim
\dot{M}_{\star}^{\rm time}$. This means that \emph{the mass reservoir
  from which stars form is not limited to the cores}: the core mass
received by the star can be replenished by accretion onto the core. In
other words, $M_{\star} > M_{\rm core}$ is entirely consistent with
the rates. For core identification schemes as adopted here, a direct
correlation of core and stellar masses is \emph{not} implied by our
observations. This is similar to models of competitive accretion,
where several stars accrete from a common environment, so that their
growth is not limited by the mass reservoir immediately surrounding them.

\section{Summary}
Although there have been many studies at high angular resolution
towards evolved high-mass objects (protostars, \HII\ regions, outflows
associated with protostars), those with the right choice of line
tracer (that trace the cold prestellar/cluster gas) with high spectral
resolution and sensitivity towards precluster gas has been few.  We
present high angular resolution observations of two cold,and high-mass
filaments G29.96e, and G35.20w, in the close vicinity of two well known
UC\HII\ regions G29.96-0.02 and W48 respectively. To our knowledge,
these are the first observations that look at the massive cold cores
with a line tracer that uniquely traces cold, and dense gas
(\DAMM). G29.96e is an IR dark cloud as seen with Spitzer at 8$\mu$m, while
except for the central object, G35.20w is IR quiet at Spitzer
24$\mu$m. The clustered nature of star formation implies that such
filaments are ideal laboratories for determining the initial
conditions of high-mass star formation.

Our findings are based on observations obtained in line (\DAMM\, \AMM\, HCO$^+$),
and continuum (3.5~mm, and 1.3~mm) with PdBI, VLA,  BIMA, SCUBA, and Spitzer. If the properties of the two filaments were to reflect the
initial conditions for forming high-mass stars, then we make the following conclusions that have to be
taken into consideration in future models of high-mass star formation;

\begin{itemize}
\item \DAMM\ is a good tracer of the entire filament seen in dust emission
  with single dish, and shows rich structure along both the
  filaments. Multiple objects were detected in \DAMM\, as well as
  \AMM.  Millimeter dust continuum detected at high S/N were detected only towards
  few of these objects. Only the brightest mm continuum core shows
  several signs of high-mass star formation, with their compact 24\,$\mu$m
  emission, high masses, densities, associated massive outflows, and methanol
  maser emission.  Therefore, gas is predominantly precluster.
\item The core temperature is always $< 22$~K as derived from our
  \AMM\ observations.
\item High deuteration seen on large scales is also observed
  on the small scales resolved by the interferometer. Such high deuteration ratios
 ([\DAMM/\AMM]$ > 6\%$), make the discoveries discussed below possible. 
  We find clear evidence of \DAMM\ being destroyed in cores showing protostellar
  activity.
\item We studied the fragmentation of these cores, and compared
  them with two local and far away low mass precluster cores. We find
  that our observations are consistent with turbulent Jeans fragmentation of massive clumps into cores (from a Jeans
 analysis).
\item  The observed velocity dispersions towards the precluster cores
  are low ($<1$~\kms), and therefore are characterized by low level of
  turbulence. 
\item The cores as well as clumps are ``over-bound/subvirial'', i.e.\ turbulence
  is too weak to support them against collapse, meaning that some
  models of monolithic cloud collapse are quantitatively inconsistent
  with data.
\item Accretion from
 the core onto a massive star which can (for observed core sizes and
 velocities) be sustained by ``fly-by accretion'' of envelope material
 onto the core, suggesting that (similar to competitive accretion
 scenarios) the mass reservoir for star formation is not necessarily
 limited to the natal core.
\end{itemize}

\acknowledgement{We thank an anonymous referee for a very thorough
  review which helped to significantly improve the paper. TP acknowledges support from the Combined Array for
  Research in Millimeter-wave Astronomy (CARMA), which is supported by
  the National Science Foundation through grant AST 05-40399. JK also thanks Di Li, his host at JPL, for making
  this research possible. This project was supported by an appointment
  of JK to the NASA Postdoctoral Program at the Jet Propulsion
  Laboratory, administered by Oak Ridge Associated Universities
  through a contract with NASA. JKs research was executed at the Jet
  Propulsion Laboratory, California Institute of Technology, under a
  contract with the National Air and Space Administration. This work is based in part on observations made with the Spitzer Space Telescope, which is operated by the Jet Propulsion Laboratory, California Institute of Technology under a contract with NASA.}
                         
\bibliographystyle{aa}
\bibliography{bib_astro}
%\listofobjects
\end{document}